\documentclass[aps,english,prb,preprint,superscriptaddress,floatfix]{revtex4}
\usepackage[?]{amsmath}
\usepackage{subfigure}
\usepackage{graphicx}

\newcommand{\unit}[1]{\,{\rm #1}}

\providecommand{\tabularnewline}{\\}

\usepackage{babel}

\begin{document}

\title{Ideal magnetohydrodynamic simulation of magnetic bubble expansion as a model for extragalactic radio lobes}

\author{Wei Liu}
\email{wliu@lanl.gov}

\affiliation{Theoretical Division, Los Alamos National Laboratory, Los Alamos, NM, USA 87545}

\author{Scott C. Hsu}

\affiliation{Physics Division, Los Alamos National Laboratory, Los Alamos, NM, USA 87545}

\author{Hui Li}

\affiliation{Theoretical Division, Los Alamos National Laboratory, Los Alamos, NM, USA 87545}

\author{Shengtai Li}

\affiliation{Theoretical Division, Los Alamos National Laboratory, Los Alamos, NM, USA 87545}

\author{Alan G. Lynn}
\affiliation{Electrical \& Computer Engineering Dept., University of New Mexico, Albuquerque, NM, USA 87131}
\date{\today}

\begin{abstract}
Nonlinear ideal magnetohydrodynamic (MHD) simulations of the
propagation and expansion of a magnetic ``bubble" plasma into a lower
density, weakly-magnetized background plasma are presented.  These
simulations mimic the geometry and parameters of the Plasma Bubble
Expansion Experiment (PBEX) [A. G. Lynn, Y. Zhang, S. C. Hsu, H. Li,
W. Liu, M. Gilmore, and C. Watts, Bull.\ Amer.\ Phys.\ Soc.~{\bf 52},
53 (2007)], which is studying magnetic bubble expansion as a model for
extra-galactic radio lobes.  The simulations predict several key
features of the bubble evolution.  First, the direction of bubble
expansion depends on the ratio of the bubble toroidal to poloidal
magnetic field, with a higher ratio leading to expansion predominantly
in the direction of propagation and a lower ratio leading to expansion
predominantly normal to the direction of propagation.  Second, an
MHD shock and a trailing slow-mode compressible MHD
wavefront are formed ahead of the bubble as it propagates into the
background plasma.  Third, the bubble expansion and propagation
develop asymmetries about its propagation axis due to reconnection
facilitated by numerical resistivity and to inhomogeneous angular
momentum transport mainly due to the background magnetic field.  These
results will help guide the initial experiments and diagnostic
measurements on PBEX.

\end{abstract}


\maketitle


\section{Introduction}
Cavities with diameters from several to tens of kiloparsecs have been
observed in the X-ray emission from  nearly two dozen galaxies, groups, and clusters.
\citep{brm04} These cavities are filled with magnetic fields and relativistic plasmas that radiate in radio
emission from ``radio lobes.''\citep{mwn00,mnw05} These observations suggest that
the X-ray cavities have formed by shoveling aside
thermal cluster plasmas by radio-emitting plasmas emanating from
galaxies. The interactions of radio-emitting outflows with X-ray emitting cluster plasmas lead to shocks, which are a candidate for heating  cluster plasmas to $\gtrsim1\unit{keV}$.\citep{mwn00,mnw05}
Past theoretical models of these systems assume that such outflows are
kinetic energy dominated (so-called kinetic energy dominated regime).
\citep{nbs88,mmi94,cd96,brm98,tjr04} However, recent observations show that both cluster and radio lobe plasmas
have appreciable magnetic energy. \citep{oek00,kdl01,fl01,chh05}
This has led to new models in which radio lobes are thought to be gigantic ``relaxed"
plasmas with kilo-to-megaparsec scale jets providing a source of
magnetic energy and helicity from the galaxy to the lobes. \citep{llf06,tx07} However, the
details of how radio lobe magnetic energy and helicity evolve and interact with the intergalactic medium are not well understood.
\citep{nm04,llf06,nll06,nll07} These details depend on underlying nonlinear plasma physics, including magnetic relaxation of radio lobe plasmas
as they expand against a background plasma while being driven by jets, heating of the lobe and background plasmas due to reconnection and shocks, and angular momentum transport within the lobe and between the lobe and background.

In order to develop further insights into extragalactic radio lobes, a laboratory
plasma experiment called the Plasma Bubble Expansion Experiment (PBEX)
\citep{lzh07} has been built to address some of the
underlying nonlinear plasma physics issues upon which leading radio lobe models are
based.  The experiment will study the related model problem of a
magnetic plasma ``bubble" relaxing and expanding into a lower pressure weakly-magnetized background plasma. A new
pulsed coaxial gun will form and inject magnetized plasma bubbles ({\em
i.e.}, the lobe) into a background plasma ({\em i.e.},
the intergalactic medium) formed by a helicon and/or hot cathode
source on the HELCAT facility.\citep{lgw08}  Experimental parameters can be adjusted so that
important dimensionless parameters, such as plasma $\beta$, are relevant to the astrophysical
context.

Numerical modeling helps guide the experiments and aids the data interpretation.
In this paper we report initial nonlinear simulation results
performed with a new three-dimensional (3D) ideal MHD package,\citep{ll03} which is a
time-explicit, compressible, ideal MHD parallel 3D code, using
high-order Godunov-type finite-volume numerical methods, in Cartesian
coordinates $(x,y,z)$. The simulations mimic PBEX and use experimentally measured or inferred parameters
(see Table~\ref{normalization}).

This paper is organized as follows. In Sec.~\ref{setup}, we outline
the problem setup including initialization of the bubble and background plasma
column. We present the simulation results in Sec.~\ref{results}, and
discussions and implications of our results for the experiment are given
in Sec.~\ref{discussions}.

\section{Problem setup}\label{setup}

\begin{figure}[!htp]
\begin{center}

  \scalebox{0.4}{\includegraphics{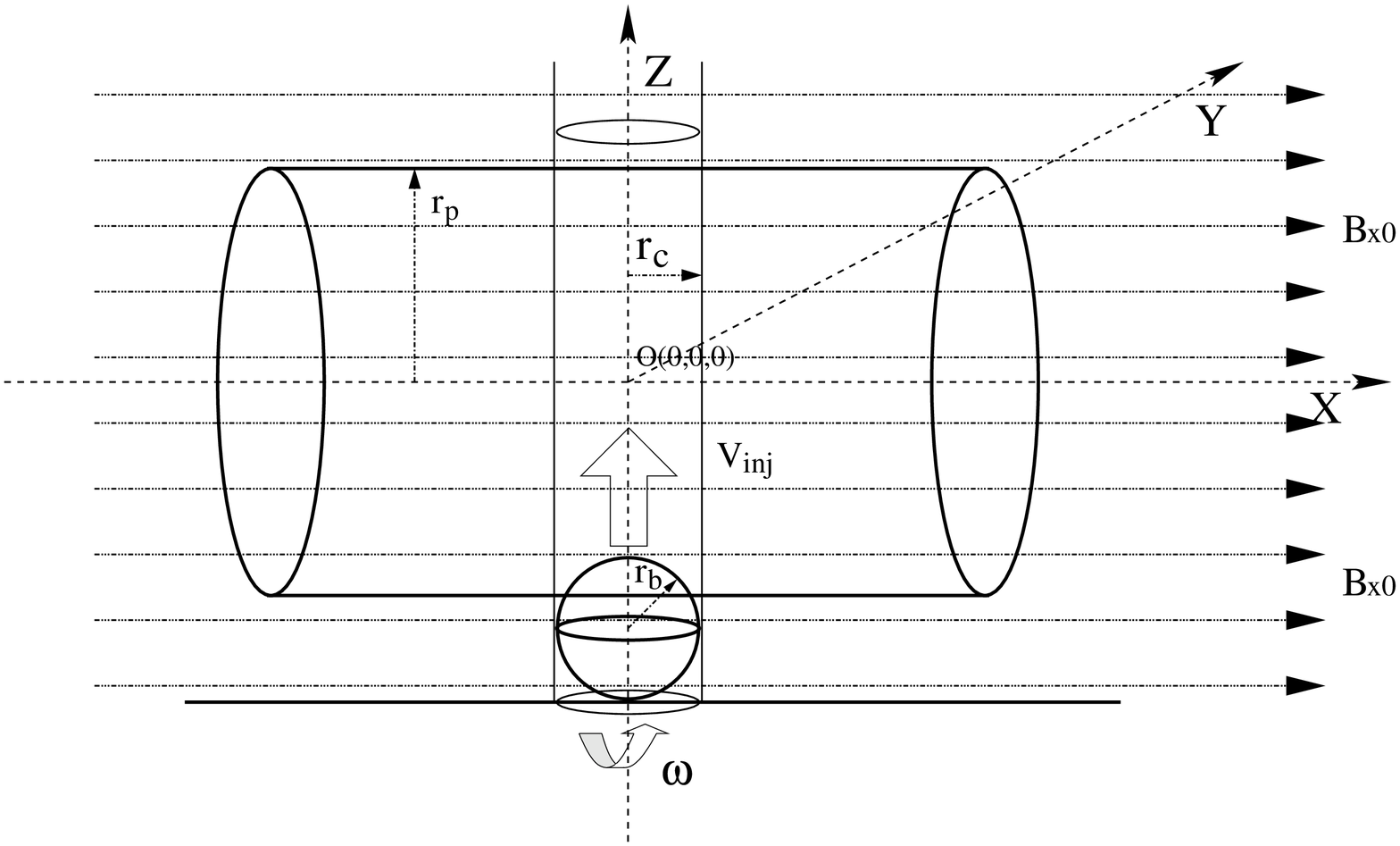}}
  \caption{\label{config}~Schematic of the simulation geometry of PBEX showing also the coordinate system. In the texts, the direction along $z-$axis is defined as axial direction. $x-y$ plane is defined as the toroidal plane while $x-z$ plane is defined as the poloidal plane.}
\end{center}
\end{figure}

In the simulations and experiment, a high density magnetized rotating
bubble plasma is injected radially into a cylindrical
plasma volume with a background magnetic field, as shown in Fig.~\ref{config}. The injected magnetic
configuration is not force-free so that Lorentz forces cause the bubble to
expand while traveling through and interacting with the background
plasma. The basic model assumptions and numerical treatments we adopt
here are essentially the same as those in \citet{llf06}
The nonlinear system of time-dependent ideal MHD
equations in 3D Cartesian coordinates $(x,y,z)$ is given here:
\begin{eqnarray}
\frac{\partial \rho}{\partial t}+\nabla\cdot(\rho\vec{v})&=&0\,,\\
\frac{\partial(\rho\vec{v})}{\partial t}+\nabla\cdot\left(\rho\vec{v}\vec{v}+(p+\frac{B^2}{2})\mathbf{I}-\vec{B}\vec{B}\right)&=&0\,,\\
\frac{\partial E}{\partial t}+\nabla\cdot\left[\left(E+p+\frac{B^2}{2}\right)\vec{v}-\vec{B}(\vec{v}\cdot\vec{B})\right]&=&0\,,\\
\frac{\partial \vec{B}}{\partial t}-\nabla\times(\vec{v}\times\vec{B})&=&0\,,
\end{eqnarray}
in which $\rho$, $p$, $\vec{v}$, $\vec{B}$ and E are the density, (gas) pressure, flow velocity, magnetic field, and total energy, respectively. $\mathbf{I}$ is the unit diagonal tensor. The total energy is $E=p/(\gamma-1)+\rho v^2/2+B^2/2$, where $\gamma=5/3$ is the ratio of the specific heats. Note that a factor of $\sqrt{4\pi}$ has been absorbed into the scaling for both the magnetic field $\vec{B}$ and current density $\vec{j}$.
More details are given in \citet{llf06} All simulations are performed on the parallel Linux clusters at Los Alamos National Laboratory. 
It should be noted that the details of effects such as reconnection and heat evolution cannot be addressed accurately due to the ideal MHD model and the use of a simplified energy equation.

Physical quantities are normalized by the characteristic system length scale $R_0$, density $\rho_0$, and velocity $C_{s0}$ based on the
measured or expected values from PBEX\@.  The normalization factors are summarized in Table~\ref{normalization}.  Normalized variables are used hereafter.

\subsection{Background plasma equilibria}

A higher pressure
magnetic plasma bubble with spherical radius $r_b=1$, centered initially at $x_b=0$, $y_b=0$ and $z_b=-7.33$, is injected along the $z$ axis into a lower pressure background
plasma with injection velocity $v_{\rm inj}$ (see Fig.~\ref{config}). The stationary background plasma is composed of a cylindrical plasma column with radius $r_p=6.67$ confined by a background magnetic field $B_{x,0}(r)$, where $r=\sqrt{y^2+z^2}$.  
Although PBEX will offer a choice of gas combinations for the bubble and background, the initial experiments will likely use argon for
both, and therefore the initial simulations are based on argon with atomic mass of $39.948$. 

Force balance of the background plasma along the $r$ direction gives
\[
p(r)+\frac{B_{x,0}(r)^{2}}{2}=\frac{B_{x,0}(r_p)^{2}}{2}\,.
\]

For $r\ge r_p$, $B_{x}(t=0)=B_{x,0}(r_p)$, where $B_{x,0}(r_p)$ is taken to be $3.65$ ($75\unit{G}$), while for $r< r_p$, $B_{x}(t=0)$ is determined by the initial pressure profile:
\[
B_{x,0}(r,t=0)=\sqrt{B_{x,0}(r_p)^{2}-2p(r,t=0)}.
\]

The background plasma number density and temperature profiles are given by the following functions:
\[
n_p(r,t=0)=1.06\exp(-\gamma_\rho r),\quad T_p(r,t=0)=2.7\exp(\gamma_T r)\,.
\]
which are a good fit to actual experimental data taken by a Langmuir probe on PBEX, where the typical $\gamma_\rho$ and $\gamma_T$ are taken to be $0.33$ and $0.14$, respectively.
Thus, the initial pressure profile of the background plasma is $p(r,t=0)=n_p T_p|_{t=0}\propto \exp[(\gamma_T-\gamma_\rho)r]$.

\subsection{Magnetized bubble plasma}

A higher pressure magnetized plasma bubble is generated and injected by a coaxial gun source. It is well established empirically in coaxial gun spheromak experiments that, under proper conditions, a spheromak ``magnetic bubble" will be formed by the gun discharge.\cite{gkb98,yb00,hb03,hb05} In the simulations reported here, the bubble structure is similar to the one given in \citet{llf06}

The number density profile of the bubble plasma with radius $r_b=1$ is given by
\[
n_b\propto r_c^2\exp[-r_c^2-(z_c-z_b)^2],
\]
up to a normalization coefficient $n_{b0}=100$ and a uniform temperature $T_{b0}=10$, where
$r_c=\sqrt{x^2+y^2}$ and $z_c=z$ (see Fig.~\ref{config}).
The density profile used here has its peak shifted from the center of the bubble, approximating a spheromak, and
is therefore different from the uniform density profile used in \citet{llf06}

The bubble magnetic field is determined by three key quantities: the length scale of the bubble magnetic field $r_B=1$, the amount of poloidal flux $\Psi_p$, and the index $\alpha$, which is the ratio of the bubble toroidal to poloidal magnetic fields. For simplicity, the bubble magnetic field $\vec{B}_{\rm bubble}$ is also assumed to be axisymmetric.  The poloidal flux function $\Psi_p$ is specified as:
\begin{equation}
\Psi_p\propto r_c^2\exp[-r_c^2-(z_c-z_b)^2]\,.
\end{equation}
The poloidal fields, up to a normalization coefficient $B_{b0}=48.7$ ($1000\unit{G}$), are:
\begin{equation}
B_{{\rm bubble},r_c}=-\frac{1}{r_c}\frac{\partial \Psi_p}{\partial z_c},\quad B_{{\rm bubble},z_c}=\frac{1}{r_c}\frac{\partial \Psi_p}{\partial r_c},
\end{equation}
while the toroidal magnetic field is
\begin{equation}
B_{{\rm bubble},\varphi_c}=\frac{\alpha \Psi_p}{r_c}=\alpha r_c\exp[-r_c^2-(z_c-z_b)^2]\,.
\end{equation}
The azimuthal component of the bubble Lorentz force is zero, but the total azimuthal Lorentz force due to the
\emph{combined} fields and currents of the bubble and the background plasma may be non-zero. 

The bubble also has uniform injection velocity $v_{\rm inj}$ and uniform rotation angular speed $\Omega=\sqrt{4\pi}V_{A,0}/r_b$, where $V_{A,0}=B_{b0}/\sqrt{4\pi\rho_{b0}}=4.87$. Please note that this is a strong rotation, possibly having strong influence on the stability of the bubble (Sec.\ref{stability}) and the expansion of the bubble in the $x$-$y$ plane (Sec.\ref{internal}).

\subsection{Computational domain}
The total computational domain is $|x|\le9$, $|y|\le9$, and $|z|\le9$, corresponding to a $(54\;\unit{cm})^3$ box in actual length units. The numerical resolution used here is $400\times400\times400$, where the grid points are assigned uniformly in the $x-$, $y-$, and $z-$directions. A cell $\delta x$ $(=\delta y=\delta z=0.045$) corresponds to $0.135\;\unit{cm}$. We use ``outflow" boundary conditions at every boundary, \emph{i.e.}, setting all values of variables in the ghost zones equal to the values in the corresponding active zones, which is the simplest approach possible. This technique is accurate for supersonic outflow but not for subsonic outflow. This simplified boundary condition limits our ability to predict the transit time, the time for the bubble to travel through the background plasma, and to study the detachment problem, \emph{i.e.}, under what conditions the bubble would separate from the wall boundary.  More accurate
boundary conditions will be implemented in future work.  Here, we focus on the interaction of the bubble plasma with the background plasma before the structures have reached the boundaries.

\section{Simulation results}\label{results}

In this section we present ideal MHD simulation results on the nonlinear evolution of a magnetic ``bubble" plasma propagating and expanding into a lower pressure background plasma.  The results are organized into three primary topics:  (1)~global evolution of the bubble-background system and interface, (2)~internal bubble evolution, and (3)~angular momentum transport both outside and inside the ``bubble."  Key findings include the formation of both an MHD shock and a reverse MHD slow-mode wavefront as a result of the bubble propagating into the background plasma, and the outward transfer of azimuthal angular momentum inside the bubble due to advection and inhomogeneous transport outside the bubble due to the background magnetic field. Please note that all physical quantities, such as the magnetic field $\vec{B}$ and flow velocity $\vec{v}$, presented in this section are the total value due to both the bubble and background plasmas.

\subsection{Global evolution of the bubble-background system and interface}

In this subsection, we examine the evolution of the global bubble-background system and the interface between the two plasmas.

\subsubsection{Bubble propagation and expansion}\label{sec:overview}
Here we discuss the time evolution of the magnetic bubble, showing selected physical quantities using 2-D $x$--$z$ slices at $y=0$. The density distributions at various times $(t=0,0.25,0.5,1.0)$ are shown in Fig.~\ref{overview} with
$\alpha=\sqrt{10}$ and injection velocity $V_{\rm inj}=0.18V_{A,0}$. In this example $\alpha=\sqrt{10}$ corresponds
to the bubble having a minimum initial Lorentz force (see discussion below). At $t=0.5$, we see that the initial peak-shifted high density magnetic bubble has been transformed into a ``crab" (due to the low $\alpha$, see discussion below), bounded by one MHD shock (see Sec.~\ref{shock}) and one reverse slow-mode compressible MHD wavefront (Sec.~\ref{wavefront}).  Low-density cavities (a factor of dozens of times of magnitudes smaller than the peak density)
exist both between the shock and wavefront and in the post-wavefront region. At $t=1.5$, the shock has reached the other side of the computation domain, while the wavefront is located at $z\sim1$. The bubble is still in the middle of the background plasma. The simulation after $t=1.5$ is not accurate due to the simplified boundary conditions.

\begin{figure}[!htp]
\begin{center}
\subfigure{\scalebox{0.4}{\includegraphics{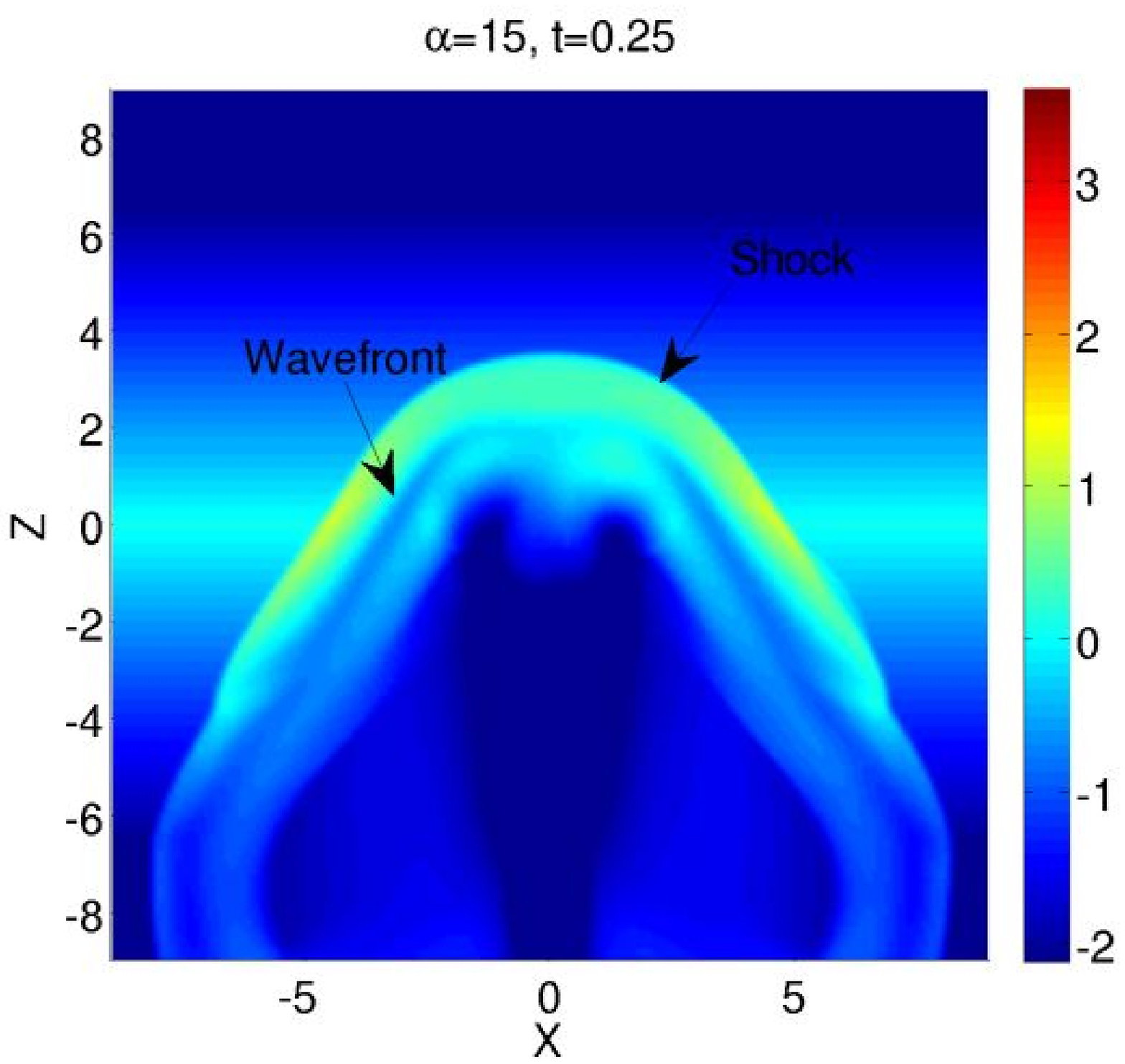}}}$\,$
\subfigure{\scalebox{0.4}{\includegraphics{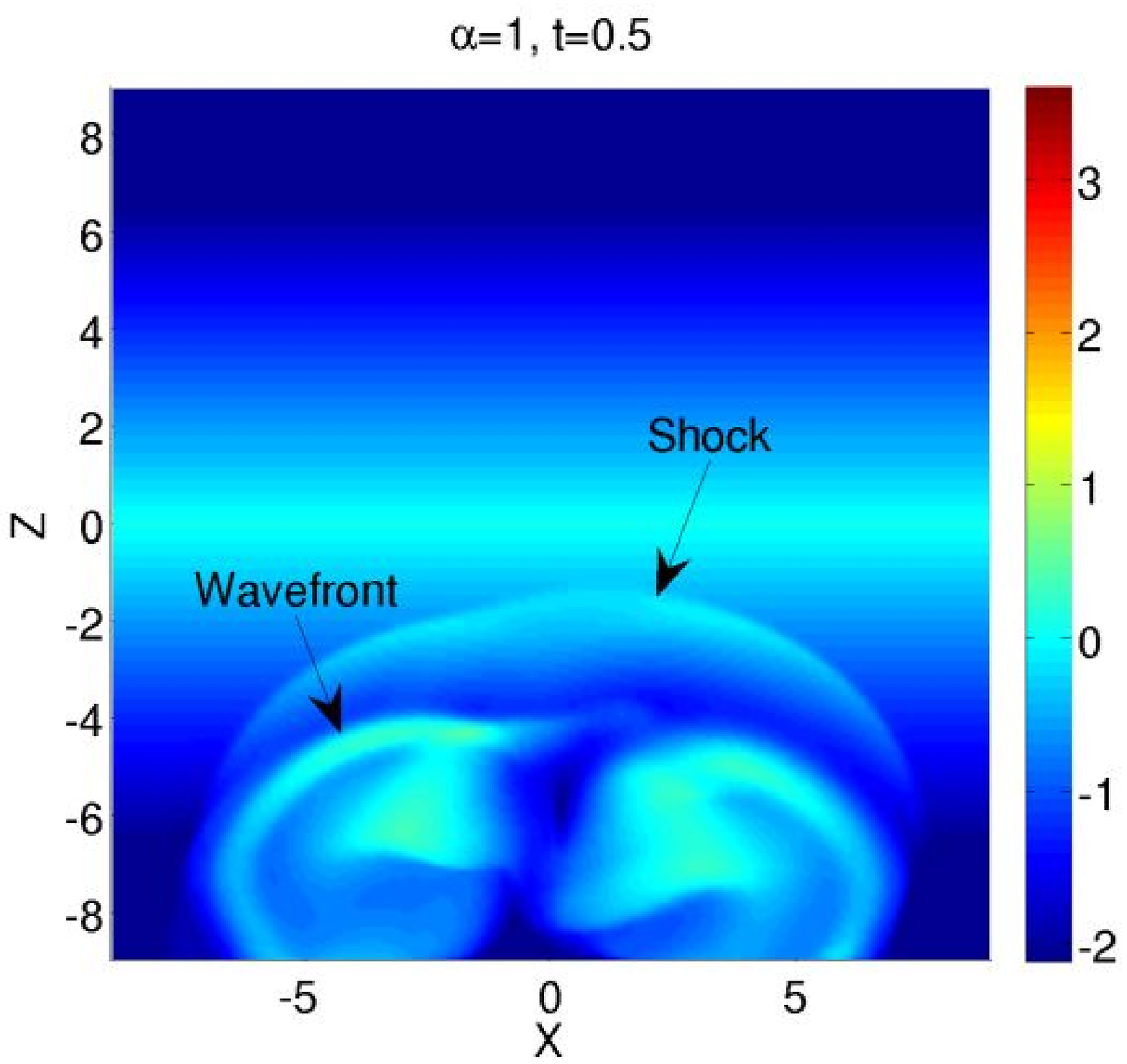}}}
\end{center}
\caption{(color) Density (natural logarithmic scale) for $\alpha=15$ (left) and $\alpha=1$ (right).
\label{fig:dependence} }
\end{figure}

The value of $\alpha$ determines the strength of the initial Lorentz force in the bubble and consequently how the bubble expands and evolves. 
We first test the influence of $\alpha$ on the bubble evolution with fixed injection speed $v_{\rm inj}$. The simulations show that the results are insensitive to $\alpha$ except for large $\alpha=15$ when the bubble expands more in the direction of the injection, leading to a growing ``mushroom"  (Fig.~\ref{fig:dependence}(left)).  With smaller $\alpha=1$, the bubble expands more transversely to the direction of injection, resulting in a growing ``crab" (Fig.~\ref{fig:dependence}(right)). This can be understood from the initial poloidal Lorentz force due to the coupling of the bubble's current and its own field (Eq.~22 and Eq.~23 of \citet{llf06}): larger $\alpha$ would have a positive axial Lorentz force but negative radial Lorentz force, resulting in collimation, while smaller $\alpha$ would have a negative axial but positive radial Lorentz force, resulting in radial expansion. In the experiment, it is expected that the ejected bubble will quickly reach a nearly force-free state. Therefore, hereafter, we assume $\alpha=\sqrt{10}$, corresponding to minimum initial bubble Lorentz force.

Given reasonably low injection velocity $(v_{\rm inj}\lesssim V_{A,0})$, there are always an MHD shock and a reverse slow-mode compressible MHD wavefront, whose structures, evolution, and propagation characteristics are essentially the same. And the injection velocity has little influence on the shock speed and wavefront speed (see Table~\ref{speed}), which implies that the shock and wavefront result from the expansion of the bubble due to the Lorentz force, rather than a ``piston effect" of the bubble propagation. However,
the ``piston" effect could become important with larger injection speeds $(v_{\rm inj}\gtrsim V_{A,0})$ (see Table~\ref{speed}), which, however, will not occur in the experiment.  For the following analysis, we will focus on the case $v_{\rm inj} = 0.18 V_{A,0}=0.88$ as being representative for the experiment.
 
\subsubsection{Identification of an MHD shock}\label{shock}
The expansion of the magnetic bubble generates a leading MHD shock and a trailing reverse slow-mode compressible MHD wavefront.  The details of these two structures are presented in this and the next subsections. The characteristics of the shock propagation are similar in the $x-$ and $z-$directions. Thus, we will present results and analysis in the $z-$direction only. 

\begin{figure}[!htp]
\subfigure{\scalebox{0.4}{\includegraphics{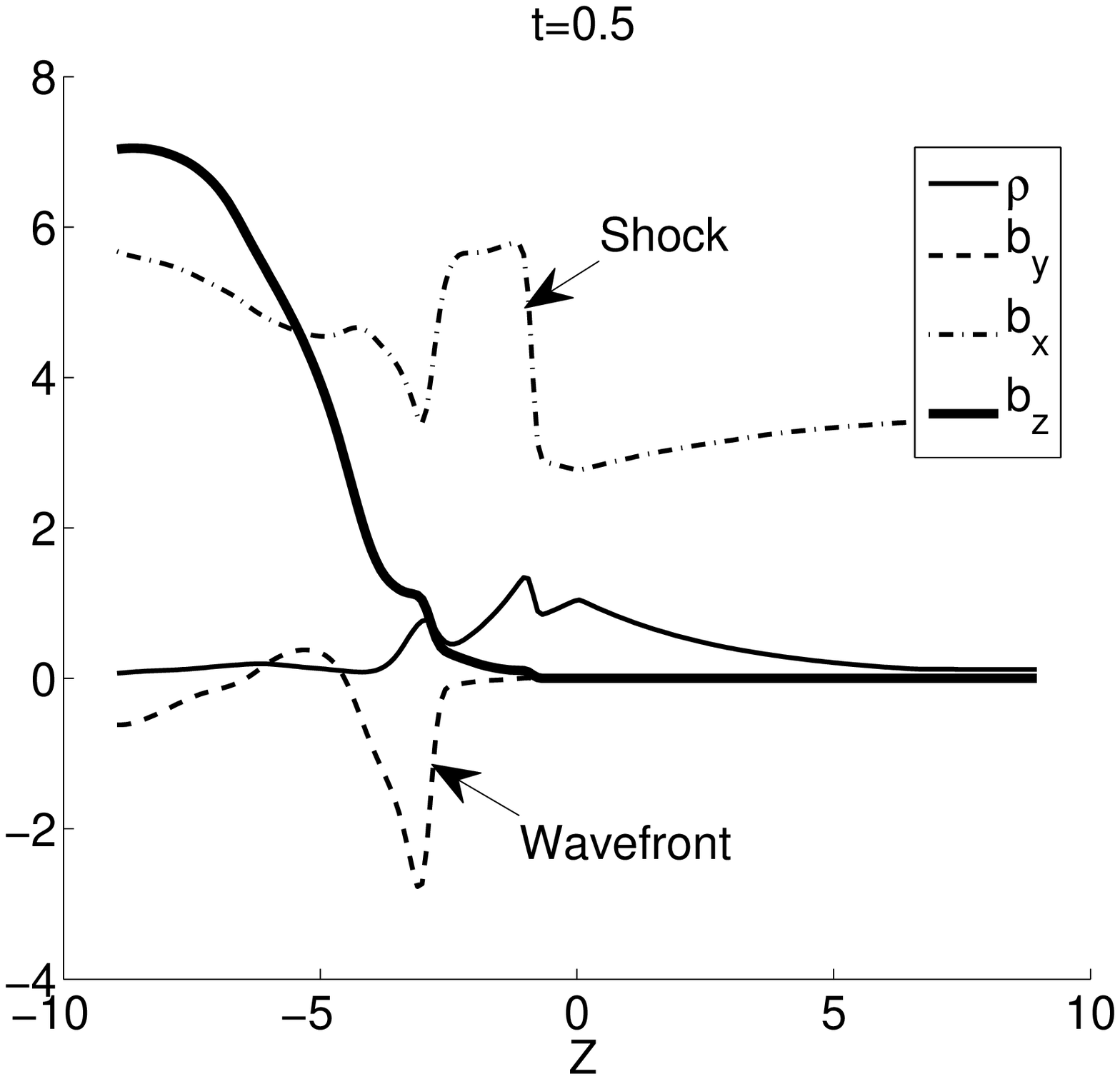}}}$\,$
\subfigure{\scalebox{0.4}{\includegraphics{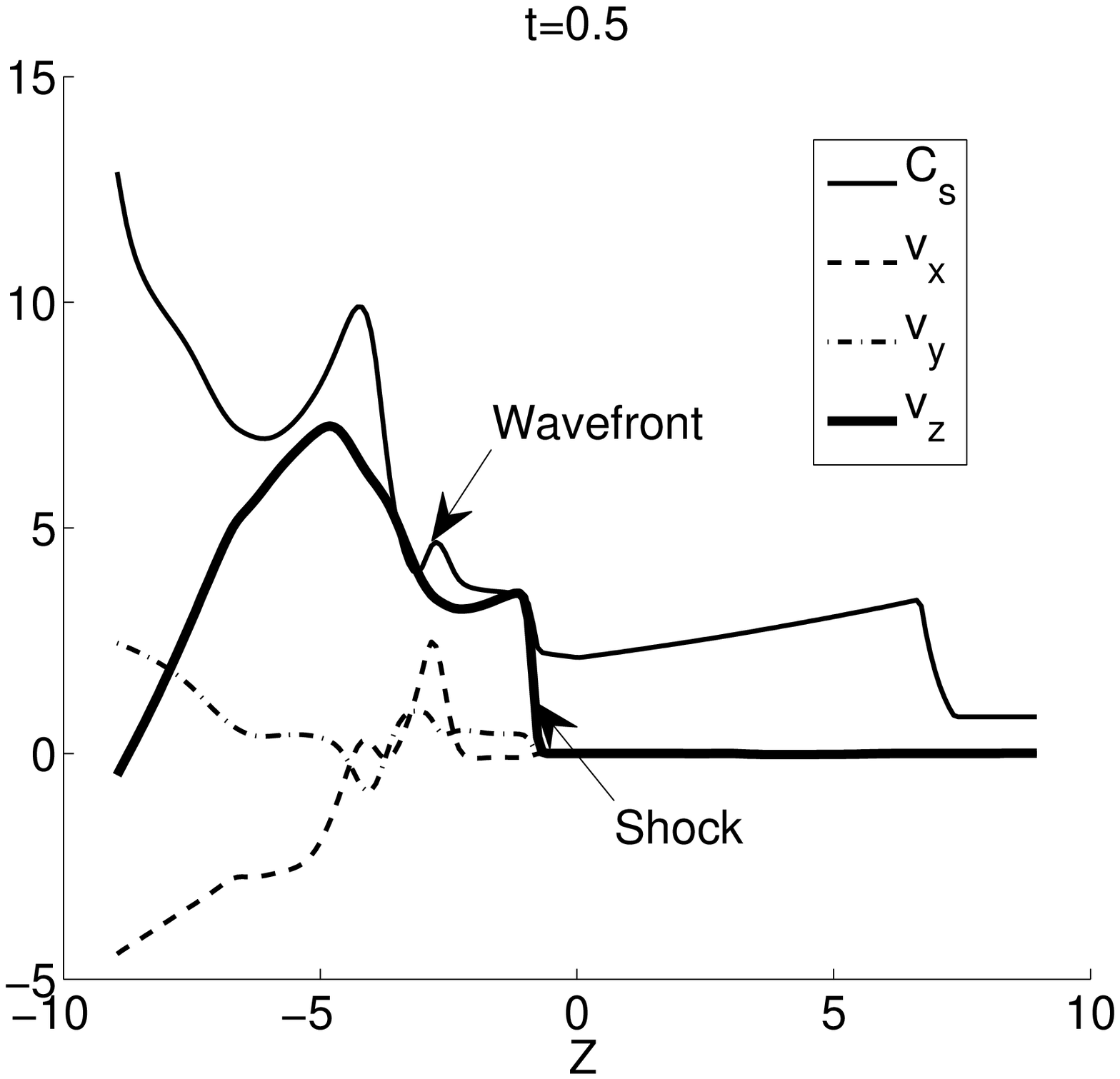}}}
\caption{\label{structure}Axial profiles of physical quantities at $(x,y)=(0,0)$ and $t=0.5$. Left:  density $\rho$ and magnetic field components $(B_x,B_y,B_z)$.  Right: sound speed $C_s$ and velocity components $(v_x,v_y,v_z)$.  \label{identification}}
\end{figure}

Figure~\ref{structure}, with $v_{\rm inj}=0.18V_{A,0}$, displays several physical quantities along a line 
with $(x,y)=(0,0)$ in the $z$-direction at $t=0.5$. Hereafter we define axial direction as the direction along $z$-axis, $x$-$y$ plane as toroidal plane and $x$-$z$ plane as poloidal plane. Several features can be identified. First, an MHD shock can be seen around $z=-1.035$ in the profiles of $\rho$, $B_x$ [see Fig.~\ref{identification}(left)] and $V_z$ [see Fig.~\ref{identification}(right)].
In the vicinity of $x\sim0$, since the magnetic field lies in the shock plane and is perpendicular to the shock normal, this shock is identified as a perpendicular shock in this region. 
This MHD shock is a fast shock whose properties are very close to an ordinary field-free shock. As shown in Fig.~\ref{structure}(left), the magnetic field components $B_y$ and $B_z$ change very little across the shock.

It is important to verify that this is indeed an MHD shock by comparing the simulation results to the
the shock jump conditions. Choosing the velocity frame so that the shock is at rest (shock velocity is $V_S$) and simplifying the notation for the problem, we represent quantities in the upstream region by a $0$ superscript and those in the downstream region by no superscript. Then there are $8$ known quantities $B_x^0$, $B_y^0=0$, $B_z^0=0$, $\rho^0$, $p^0$, $V_x^0=0$, $V_y^0=0$ and $V_z^0=-V_S$. There are $8$ unknown quantities in the downstream region, $\rho$, $p$, $V_x$, $V_y$, $V_z$, $B_x$, $B_y$ and $B_z$. Thus we need $8$ conditions to specify them. 
We consider the 1-D ideal MHD shock jump conditions for simplicity. The equations are as follows:
\begin{align}
\rho V_z  =  \rho^0 V_z^0\, , \\
\rho V_z^2+p+\frac{B_x^2}{2}+ \frac{B_y^2}{2} =   \rho^0 V_z^{02}+p^0+\frac{B_x^{02}}{2}\, ,\\
\rho V_xV_z- B_xB_z  =  0\, ,\\
\rho V_yV_z- B_yB_z  =  0\, ,\\
\rho V_z \frac{1}{2}(V_x^2+V_y^2+V_z^2)+\frac{\gamma}{\gamma-1}pV_z-&\nonumber\\
&V_xB_xB_z-V_yB_yB_z+B_x^2V_z+B_y^2V_z &\nonumber\\
&=\frac{1}{2}\rho^0V_z^{03}+\frac{\gamma}{\gamma-1}p^0V_z^0+V_z^0B_x^{02}\, ,\\
B_z  =  B_z^0\, ,\\
V_zB_y-B_zV_y=0\, ,\\
V_zB_x-B_zV_x=V_z^0B_x^0\, .
\end{align}
A {\it MATLAB} code was used to solve this nonlinear system of equations, given the values in the upstream region: $\rho^0$, $p^0$, $V_z^0$ and $B_x^0$. The results of $V_x$, $V_z$, $B_x$ and $B_y$ matches pretty well (see
Table~\ref{jump}). The nonzero simulation values of $V_y$ and $B_y$ result from 3-D effects. The relatively large differences seen in the values of $\rho$ and $p$ are possibly due to nonzero numerical diffusion in the simulations.

\subsubsection{Reverse slow-mode compressible MHD wavefront}\label{wavefront}

\begin{figure}[!htp]
\begin{center}
\subfigure{\scalebox{0.4}{\includegraphics{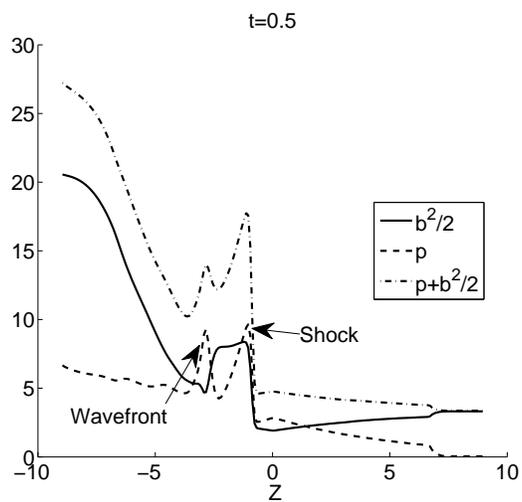}}}
\end{center}
\caption{Axial pressure profiles at $(x,y)=(0,0)$ and $t=0.5$. \label{reverse} }
\end{figure}

There is an MHD wavefront at $z=-2.835$, as seen in both panels of Fig.~\ref{structure}, where $B_x$ and $B_y$ have their local minimum, and $\rho$, $C_s$ and $v_x$ have their local maxima. The nature of this MHD wavefront can be identified by plotting the axial pressure profiles along the line $(x,y)=(0,0)$ at $t=0.5$, as shown
in Fig.~\ref{reverse}. A transition occurs near $z=-2.835$, where an increase in gas pressure $p$ is accompanied by a decrease in magnetic pressure $p_m=B^2/2$.  The transition is identified as a reverse slow-mode compressible MHD wavefront. Between the shock ($z=-1.035$) and the wavefront $(z=-2.835$), the magnetic field lines are compressed and some thermal energy has been converted into magnetic energy. Therefore the gas pressure has an abrupt decrease while the magnetic pressure increases rapidly between the shock and wavefront (see Fig.~\ref{reverse}).

\begin{figure}[!htp]
\subfigure{\scalebox{0.4}{\includegraphics{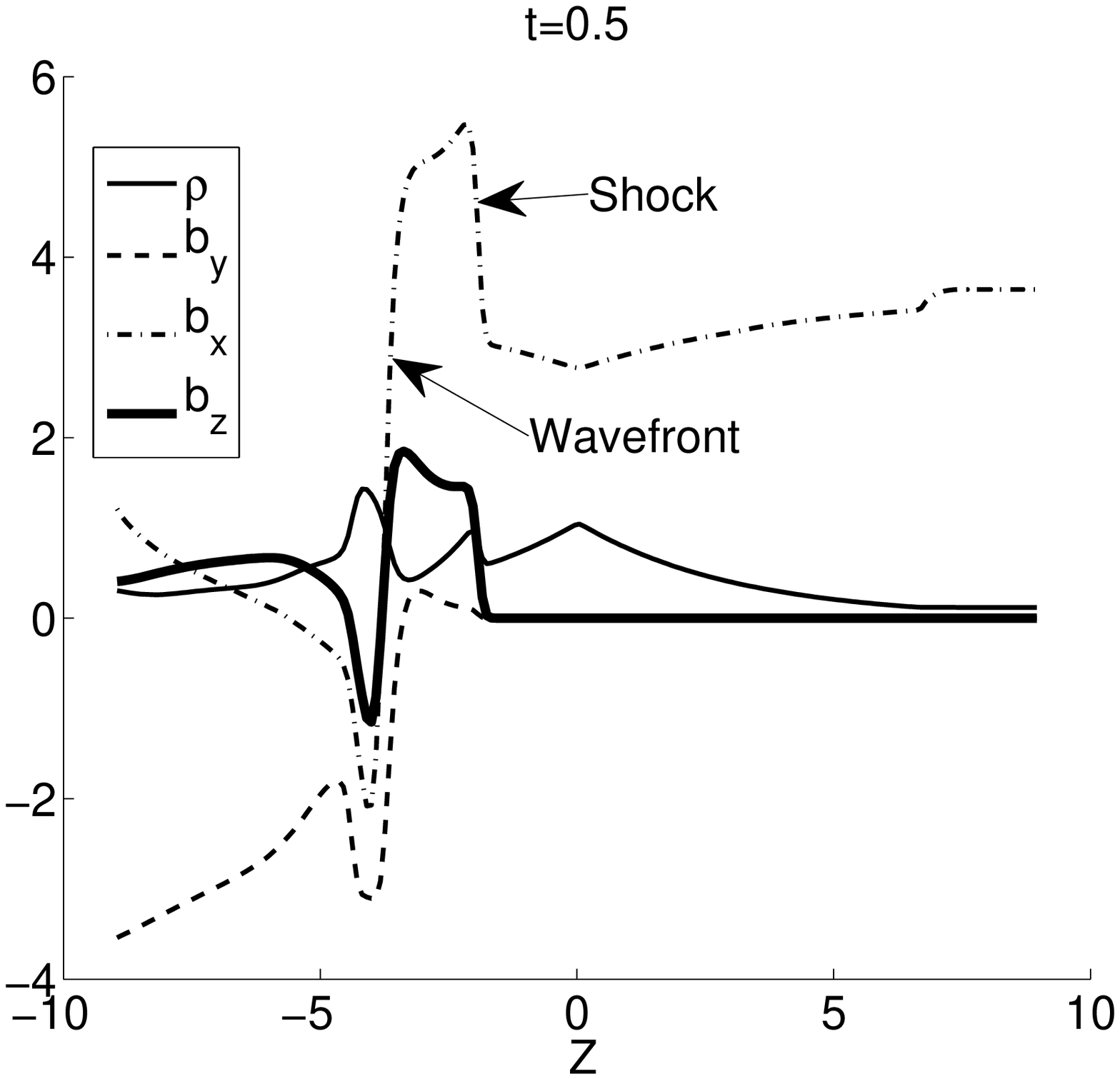}}}$\,$
\subfigure{\scalebox{0.4}{\includegraphics{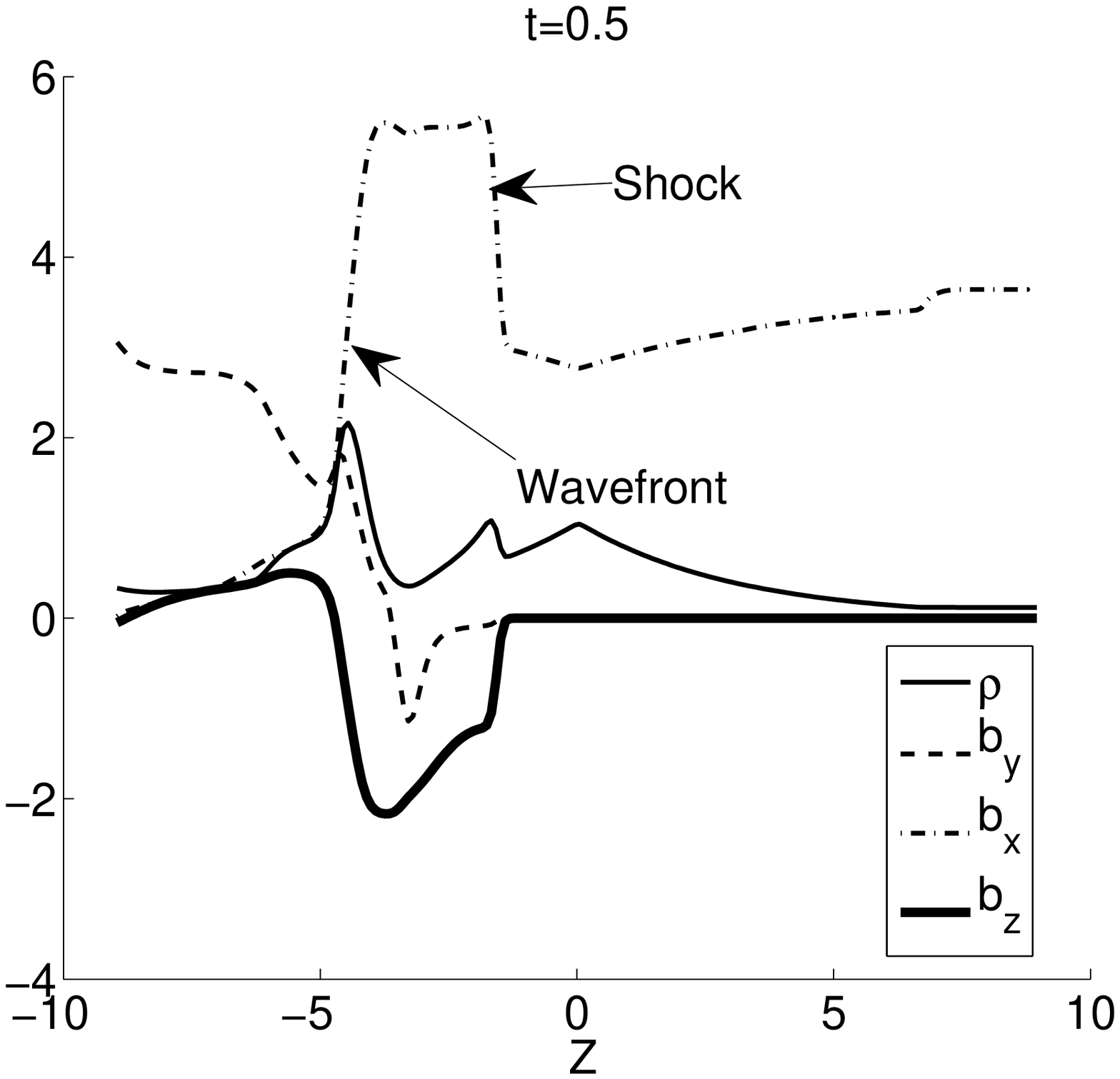}}}
\caption{Axial profiles of density $\rho$ and magnetic field components $(B_x,B_y,B_z)$ at $t=0.5$. Left: $(x,y)=(-4,0)$.  Right: $(x,y)=(4,0)$. \label{reconnection} }
\end{figure}

\begin{figure}[!htp]
\subfigure{\scalebox{0.4}{\includegraphics{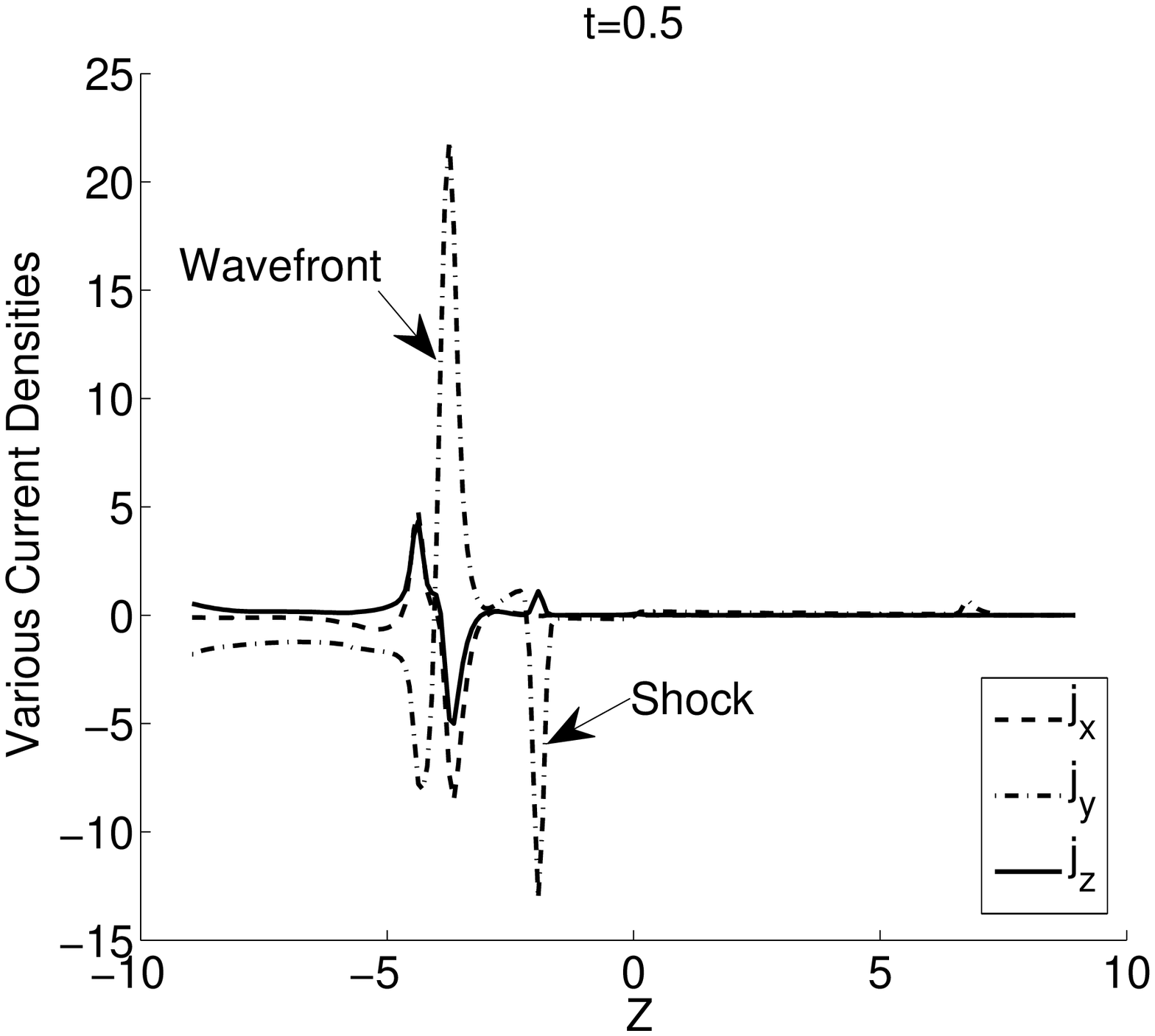}}}$\,$
\subfigure{\scalebox{0.4}{\includegraphics{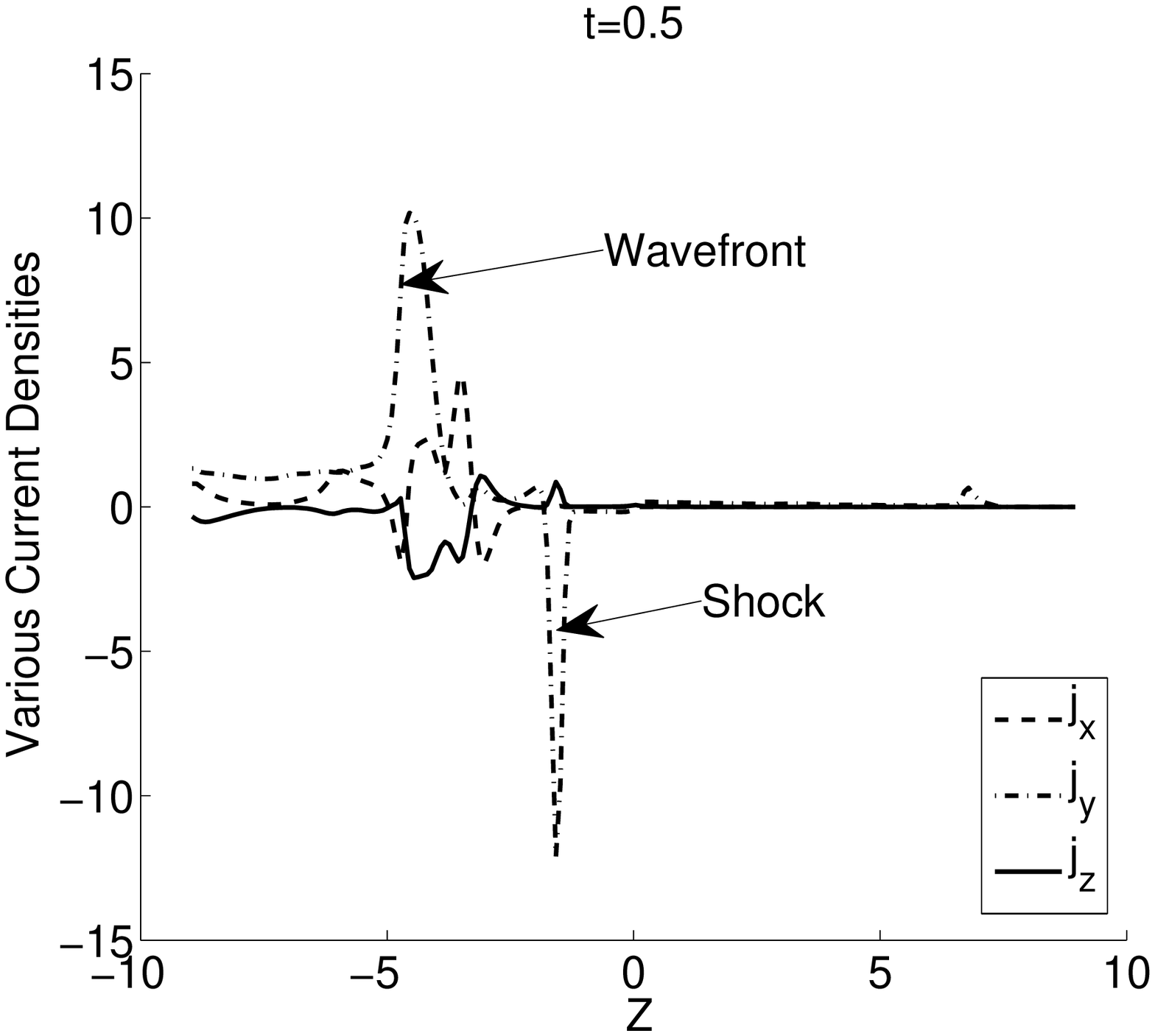}}}
\caption{Axial profiles of current density components $(j_x,j_y,j_z)$ at $t=0.5$. Left: $(x,y)=(-4,0)$.
Right: $(x,y)=(4,0)$. \label{current_z} }
\end{figure}

From Fig.~\ref{reconnection}, it can be seen that anti-parallel reconnection of the $B_x$ component occurs around
$(x,z)=(-5,-4)$, facilitated by numerical diffusion although the simulation is performed with an ideal MHD code. It is possible that $B_y$ component reconnection takes place around $(x,z)=(-4,-4)$, where the bubble field is not exactly 
anti-parallel to the background field (more details in Sec.~\ref{balance}).  The reconnection is driven by magnetic field line compression due to the reverse slow-mode compressible MHD wavefront. From Fig.~\ref{current_z}, two strong toroidal current sheets are observed at the reconnection layer and the MHD shock. These are also due to the compression of the magnetic field lines due to the wavefront and shock, respectively. The reconnection and the shock/wavefront convert normal velocity into tangent velocity and convert kinetic energy into thermal energy. Because this is an ideal MHD simulation, the details of the reconnection are not expected to be accurate. We are only interested here in the qualitative effects of the bubble evolution due to reconnection.

\subsubsection{Force in the $z$-direction}\label{balance}
\begin{figure}[!htp]
\begin{center}
\subfigure{\scalebox{0.4}{\includegraphics{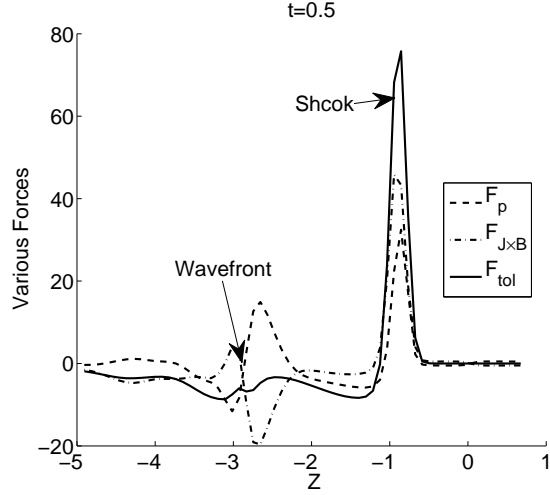}}}
\end{center}
\caption{Axial profiles of various forces at $(x,y)=(0,0)$ and $t=0.5$. $F_p$, the pressure gradient; $F_{\rm J\times B}$, Lorentz force; $F_{\rm tol}=F_p+F_{\rm J\times B}$. \label{forces} }
\end{figure}

The evolution and propagation of the magnetic bubble can be further understood by examining the various forces along
$(x,y)=(0,0)$ at $t=0.5$, which are displayed in Fig.~\ref{forces}. The MHD shock breaks the initial background equilibrium. The passage of the shock wave heats the gas and alters its pressure gradient. The axial flow is pushed forward by both the gas pressure gradient and Lorentz force at the MHD shock while it is dragged back behind the shock, resulting in an axial deceleration of the gas in the postshock region. Although the Lorentz force tries to accelerate the axial flow at the MHD wavefront, the gas pressure holds it back. Therefore the MHD shock will be driven forward and eventually separated from the wavefront, which leads to a cavity of depleted density between the shock and the wavefront.

\begin{figure}[!htp]
\begin{center}
\subfigure{\scalebox{0.4}{\includegraphics{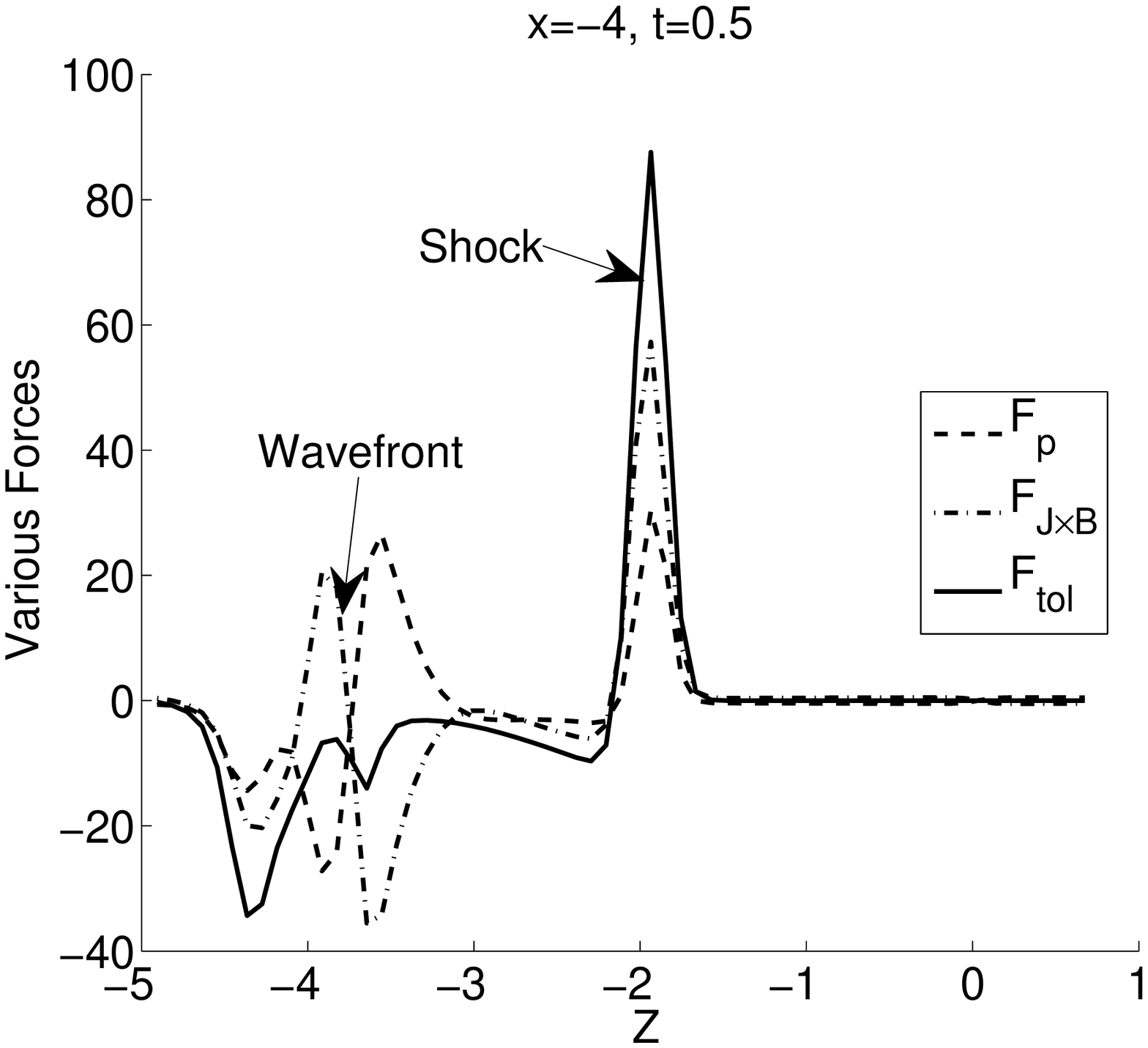}}}$\,$
\subfigure{\scalebox{0.4}{\includegraphics{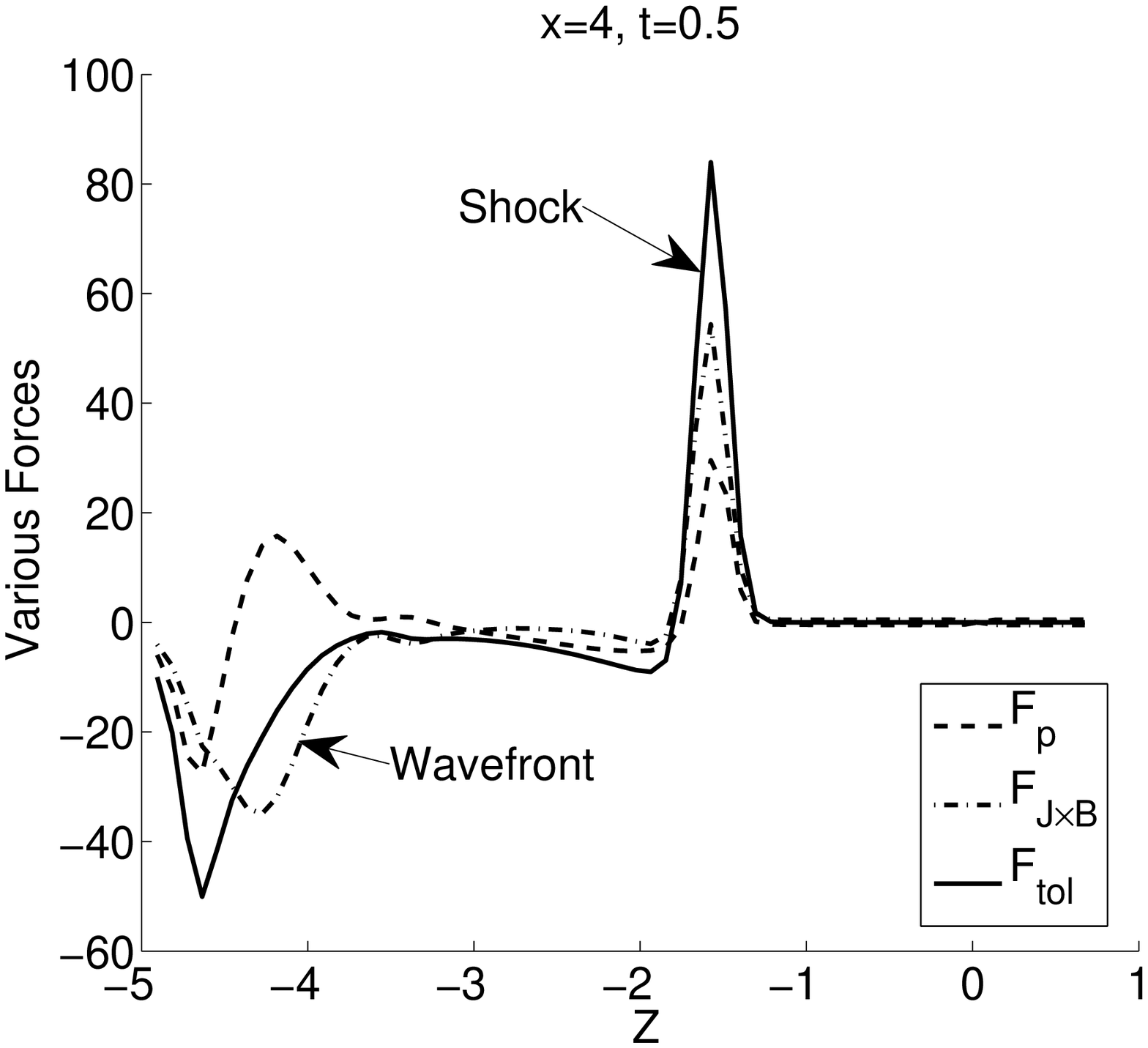}}}
\end{center}
\caption{Axial profiles of various forces at $t=0.5$. $F_p$, the pressure gradient; $F_{\rm J\times B}$, Lorentz force; $F_{\rm tol}=F_p+F_{\rm J\times B}$. Left: $(x,y)=(-4,0)$. Right: $(x,y)=(4,0)$. \label{forces2} }
\end{figure}

Figure~\ref{forces2} displays axial profiles of various forces along $(x,y)=(-4,0)$ and $(x,y)=(4,0)$ at $t=0.5$. The locations are chosen to be where anti-parallel reconnection (left panel of Fig.~\ref{forces2}) occurs and its reflection about $z=0$ (right panel of Fig.~\ref{forces2}). This figure clearly shows the difference between the two locations: the Lorentz force changes sign (similar to Fig.~\ref{forces}) on the left hand side while keeping the same sign (negative) on the right hand side, which is consistent with reconnection happening on the left hand side and not 
on the right hand side. At both $x$ locations, the toroidal current densities $j_y$ are much larger than the poloidal current densities; however, the toroidal field component $B_y$ is close to zero. Therefore $|j_yB_x|$ is much bigger than $|j_xB_y|$. Anti-parallel field lines on the left hand side result in a sign-change of the axial Lorentz force, while the Lorentz force on the right hand side does not change sign since $B_x$ does not change sign there. The sign change of the Lorentz force is necessary for reconnection since this is the driving force to pull the field lines from either side of the current sheet together to reconnect. Also, the total force is more negative on the right hand side compared to the left hand side. This means that the axial flow in the right hand side is slowed down more quickly than the left hand side, which leads to the asymmetry of the shock propagation across the $x$ axis. 

\begin{figure}[!htp]
\begin{center}
\subfigure{\scalebox{0.4}{\includegraphics{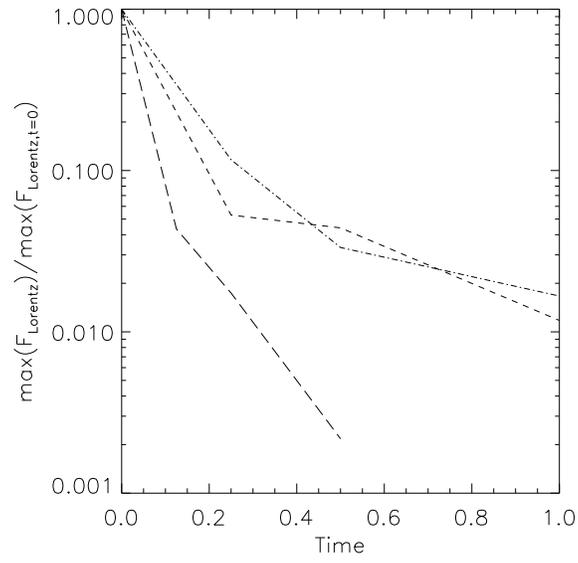}}}
\end{center}
\caption{Maximum absolute value of the axial Lorentz force over the initial value inside the bubble versus time. 
(Dash line: $\alpha=1$; dash dot line: $\alpha=\sqrt{10}$; long dash line: $\alpha=15$.)
\label{lorentz} }
\end{figure}

The magnetic bubble evolves into a nearly quasi-force-free state (see Fig.~\ref{lorentz}), with a Lorentz force that
scales in time roughly as:
\[
\frac{F_{\rm Lorentz}(t)}{F_{\rm Lorentz}|_{t=0}}\equiv\exp(-\frac{t}{\tau_{\rm relaxation}}). 
\]
The time scale of this relaxation $\tau_{\rm relaxation}$ is dependent on the value of $\alpha$, which determines the amplitude of the initial Lorentz force. Larger initial Lorentz force leads to quicker relaxation.  For 
$\alpha=1$, $\tau_{\rm relaxation}=0.379$; for $\alpha=\sqrt{10}$, $\tau_{\rm relaxation}=0.386$; and for
$\alpha=15$, $\tau_{\rm relaxation}=0.120$.
The relaxation happens on the order of the Sweet-Parker reconnection time $\tau_{\rm SP}=\sqrt{\tau_{a}\tau_{\rm res}}$
(see definitions and estimates of $\tau_{a}$ and $\tau_{\rm res}$ in Sec.~\ref{discussions}).
For example, $\tau_{\rm SP} \approx 0.3$ for $\alpha=\sqrt{10}$, which is similar to the
Lorentz force relaxation time of 0.386.  However, it should be noted that the Lorentz force in the shock/wavefront is always significant (Fig.~\ref{forces}).

\subsection{Internal bubble evolution}

In this subsection, we examine the evolution and properties of the bubble itself, including a simple kink stability analysis.

\subsubsection{Bubble density, velocity and magnetic field evolution}\label{internal}

\begin{figure}[!htp]
\begin{center}
\subfigure{\scalebox{0.4}{\includegraphics{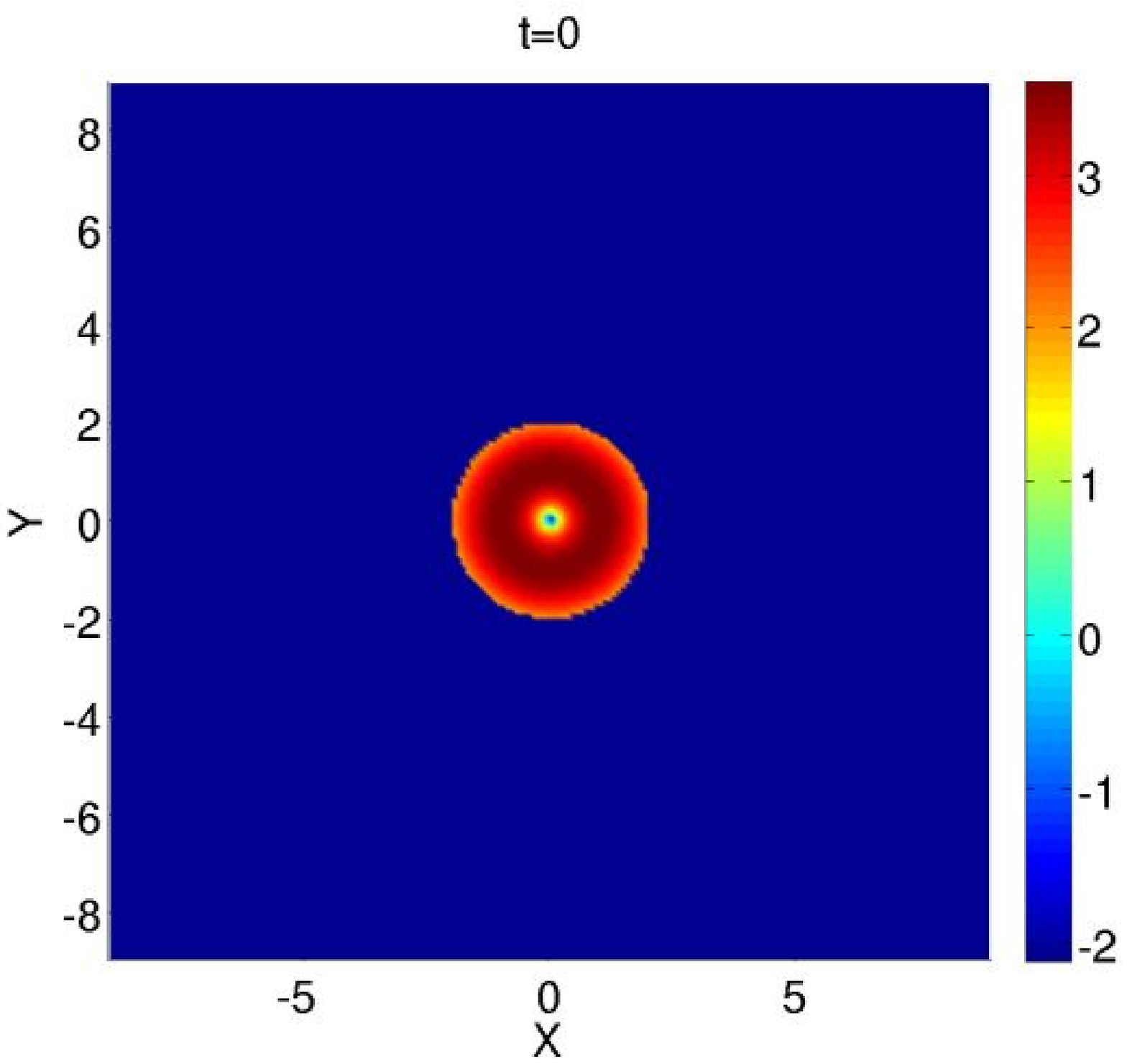}}}$\,$
\subfigure{\scalebox{0.4}{\includegraphics{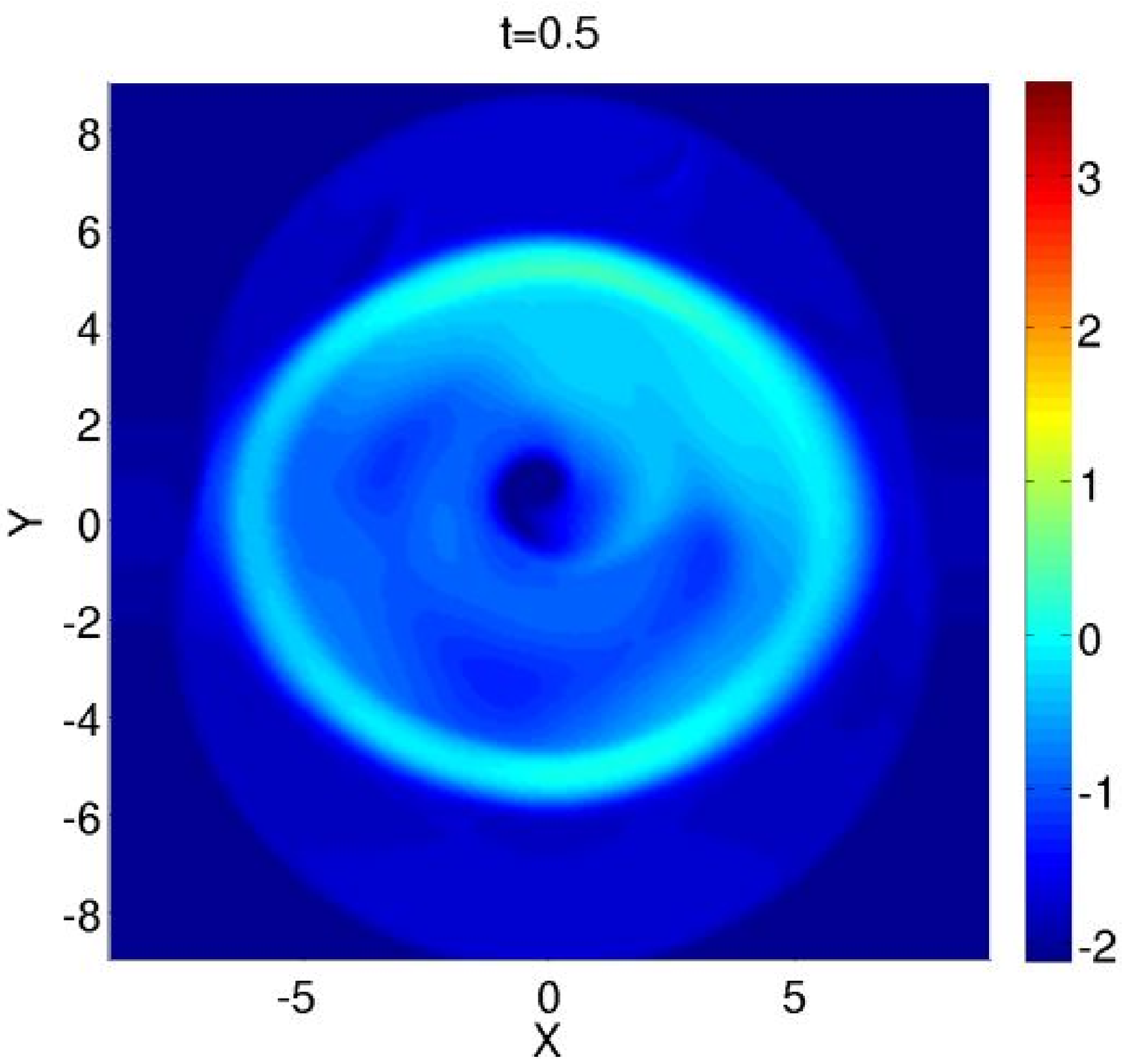}}}
\end{center}
\caption{(color) Contour plot of density $\rho$ (natural logarithmic scale) in the $x$-$y$ plane at different times. Left: $t=0$, $z=-7.5$.  Right: $t=0.5$, $z=-6$. \label{rho} }
\end{figure}

\begin{figure}[!htp]
\begin{center}
\subfigure{\scalebox{0.4}{\includegraphics{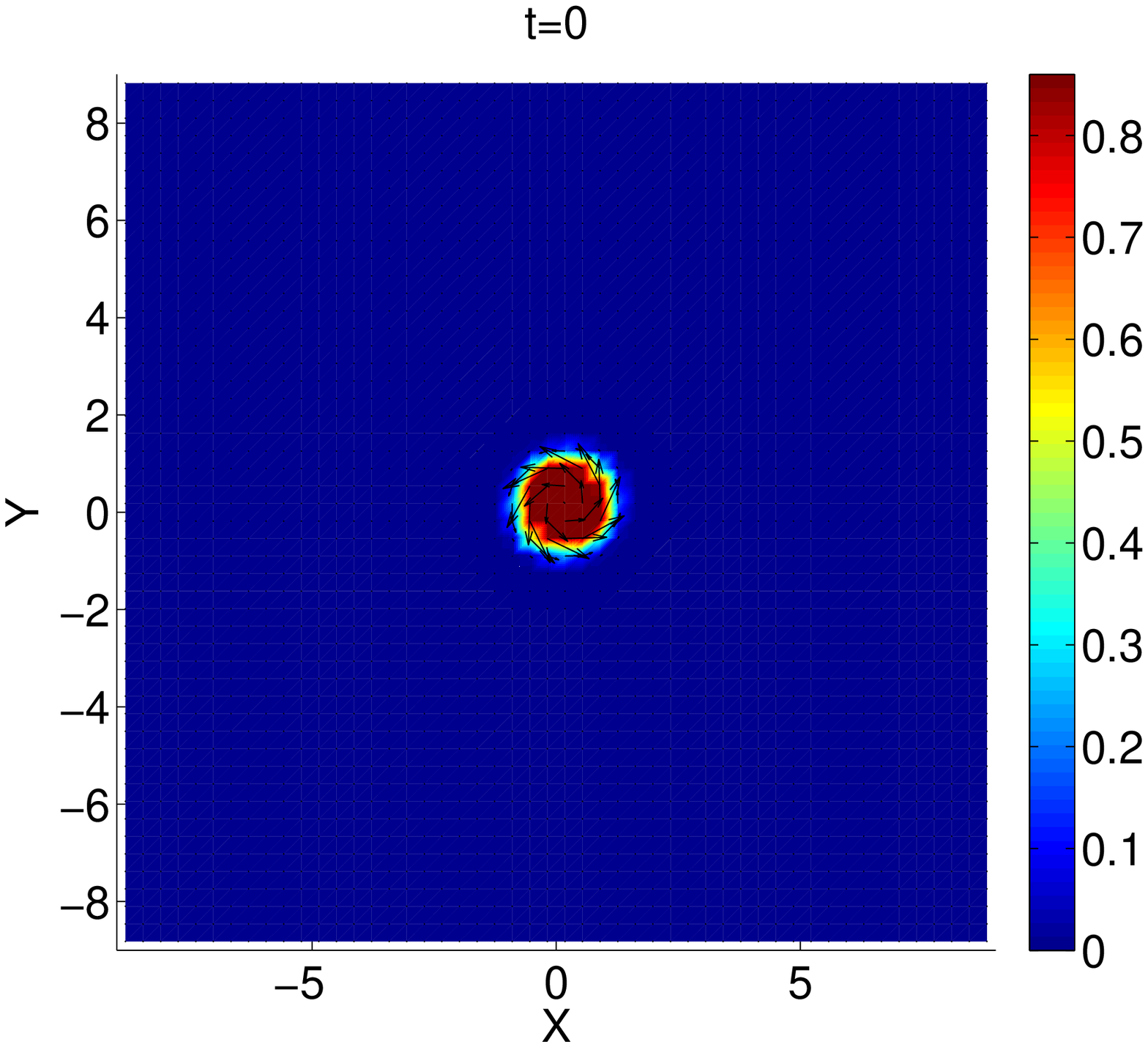}}}$\,$
\subfigure{\scalebox{0.4}{\includegraphics{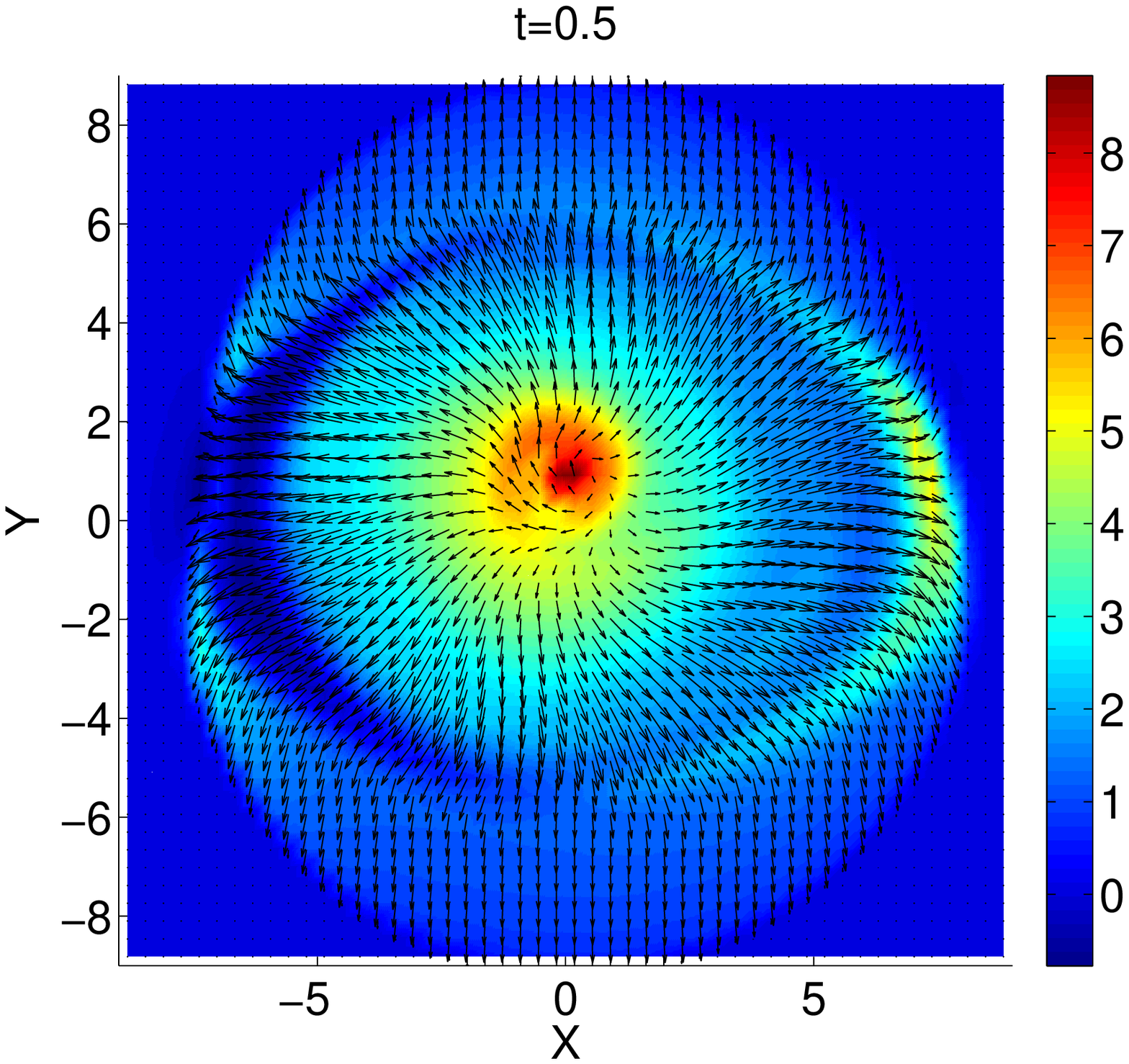}}}
\end{center}
\caption{(color) Vector plot of flow velocity $\vec{v}$ at $t=0.5$ in the $x$-$y$ plane. Arrows: flow velocity components $v_{x}$ and $v_y$; color: flow velocity $v_z$. Left: $t=0$, $z=-7.5$.  Right: $t=0.5$, $z=-6$. \label{velocity} }
\end{figure}

\begin{figure}[!htp]
\begin{center}
\subfigure{\scalebox{0.4}{\includegraphics{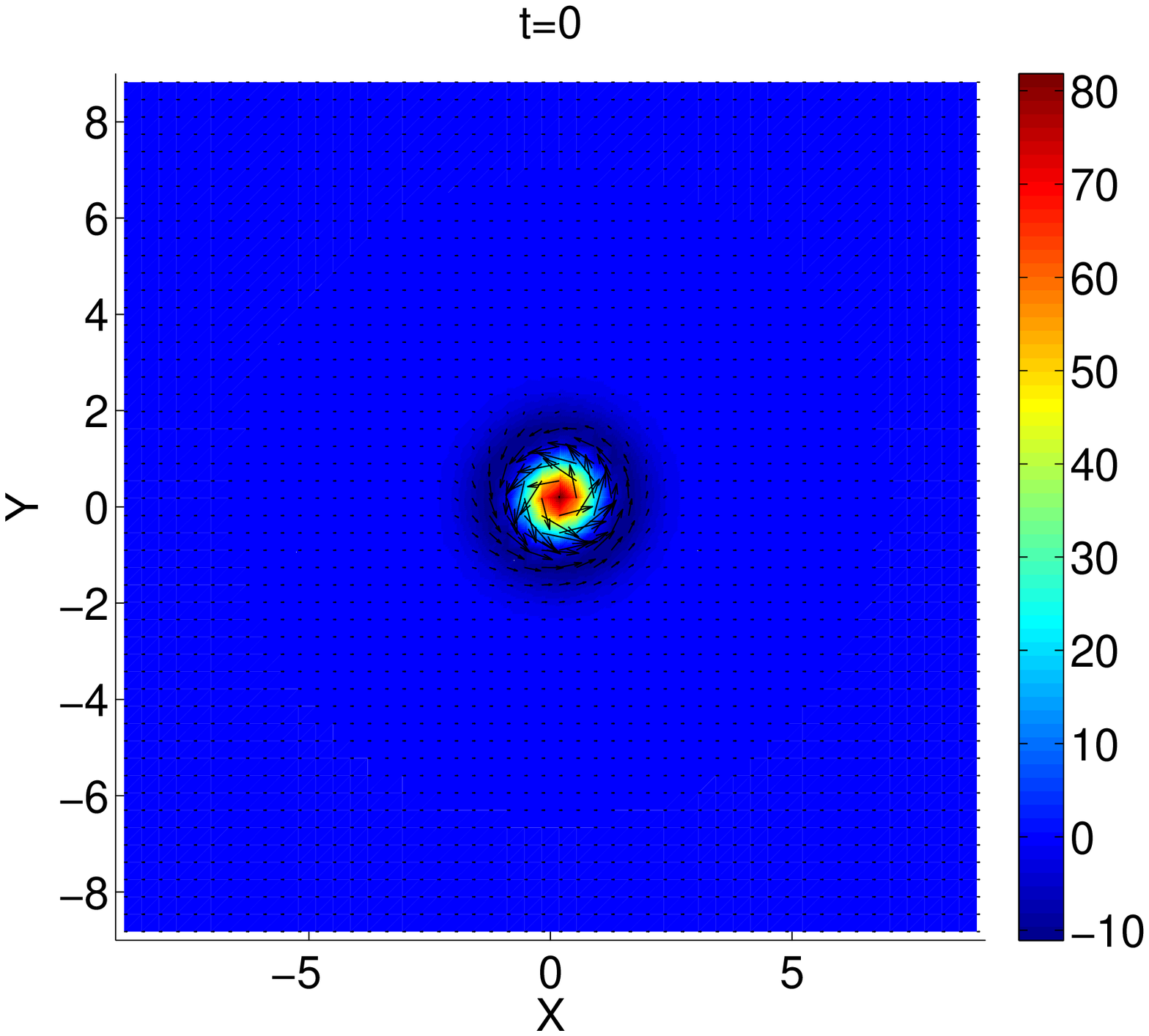}}}$\,$
\subfigure{\scalebox{0.4}{\includegraphics{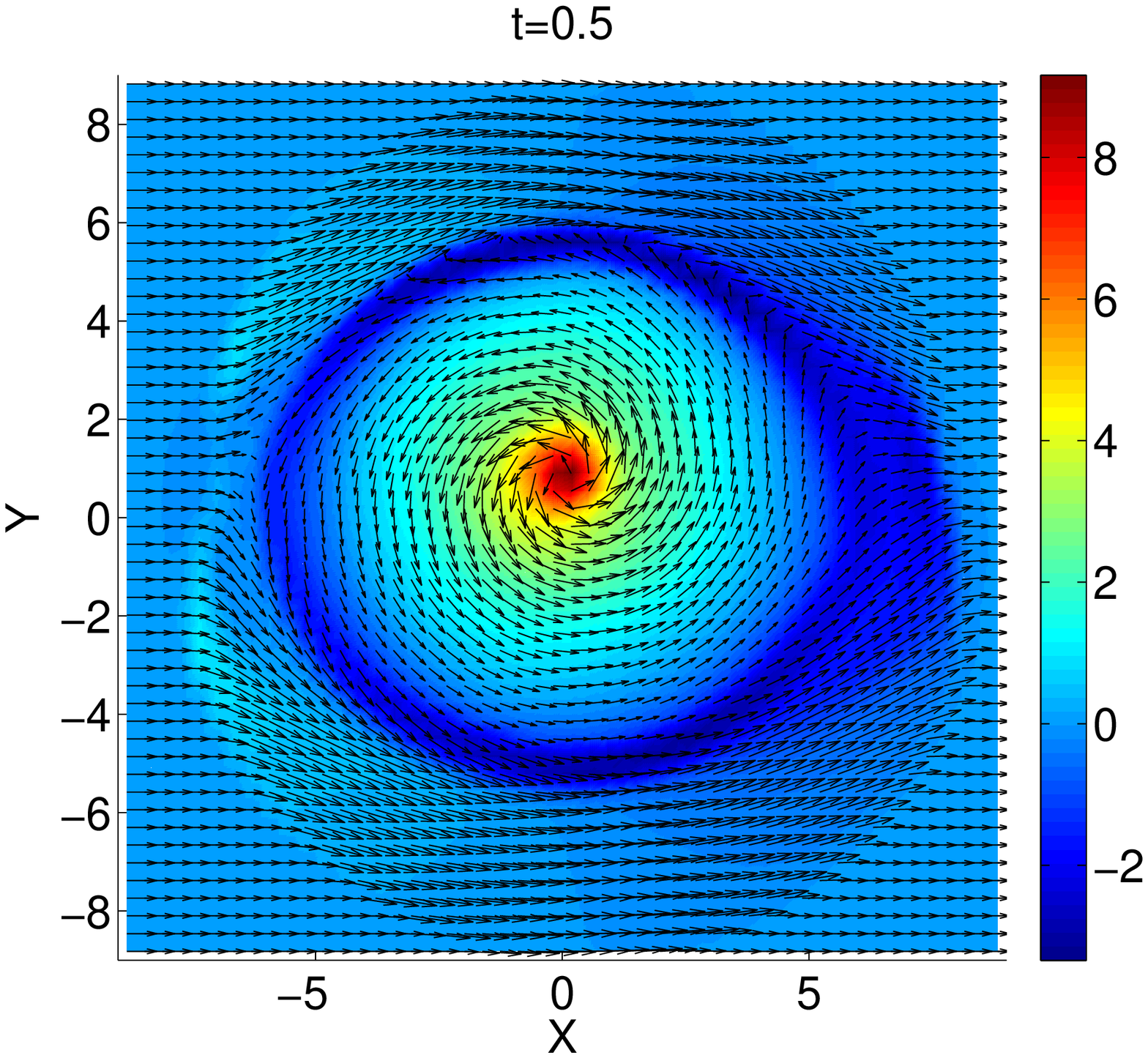}}}
\end{center}
\caption{~(color) Vector plot of magnetic field $\vec{B}$ at $t=0.5$ in the $x$-$y$ plane. Arrows: magnetic fields $B_{x}$ and $B_y$; color: magnetic field $B_z$. Left: $t=0$, $z=-7.5$. Right: $t=0.5$, $z=-6$. \label{field_xy} }
\end{figure}

Density (Fig.~\ref{rho}) and fluid velocity vector (Fig.~\ref{velocity}) plots in the $x-y$ plane at different times
both demonstrate that the initial fast-rotating spheromak-like magnetic bubble evolves into a much larger slow-rotating, fast-expanding elliptical structure with maximum density reduced by $20$ times, while the density at the wavefront and the shock increases by $8$ and $2$ times, respectively. The center of the bubble shifts away from the original propagation axis with $(x,y)=(0,0)$. Figure~\ref{field_xy} displays the vector magnetic field plots in the $x-y$ plane at different times. From the figure, we can see that the bubble field still keeps a spheromak-like configuration and the expansion of the bubble pushes away the background fields, thus compressing them.  Some reconnection happens in the region $y>0$ because the bubble toroidal field component is opposite to the background field there. These figures show that the background field breaks the symmetry of the system, which is consistent with the results of Sec.~\ref{balance} and Sec.~\ref{transport}. 

It is worth noting that the initial expansion of the bubble in $x$-$y$ plane results from the non-free initial Lorentz force as well as the centrifugal force due to the strong initial rotation of the bubble, although the latter quickly slows down to a small value because of the conservation of angular momentum associated with the initial quick expansion and the possible Kelvin-Holmhotz instability associated with the initial strong toroidal velocity shear.

\subsubsection{Bubble stability}\label{stability}

\begin{figure}[!htp]
\begin{center}
\subfigure{\scalebox{0.4}{\includegraphics{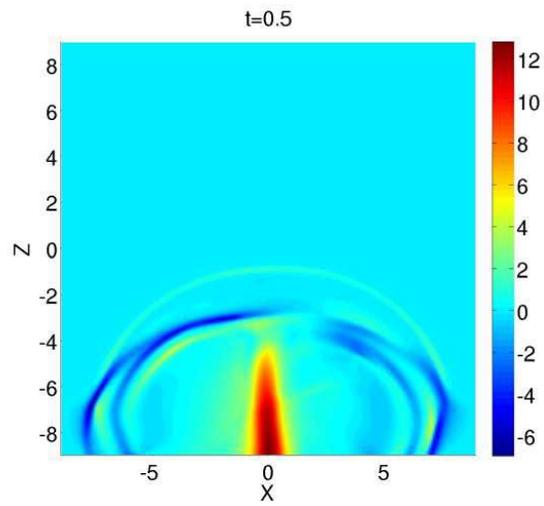}}}
\end{center}
\caption{(color) Axial current density $j_z$ at $t=0.5$ in the $x$-$z$ plane at $y=0$.
\label{jz} }
\end{figure}

\begin{figure}[!htp]
\begin{center}
\subfigure{\scalebox{0.4}{\includegraphics{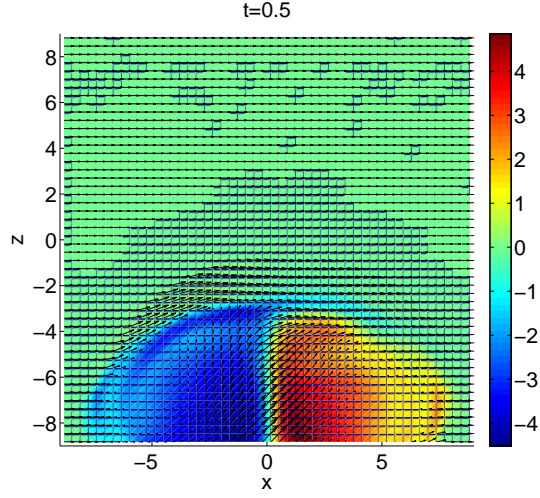}}}
\end{center}
\caption{(color) Vector plot of magnetic fields $\vec{B}$ at $t=0.5$ in the $x$-$z$ plane at $y=0$. Arrows: poloidal magnetic fields $B_{x}$ and $B_z$.  Color: toroidal magnetic field $B_y$. \label{field} }
\end{figure}

Spheromak-like bubble plasmas are subject to current-driven kink instabilities.
Figure~\ref{jz} shows a snapshot of the axial current density $j_z$ at $t=0.5$. The axial current flow follows a semi-closed (it will close outside the out-flowing boundary) circulating path, flowing along the central axis (the ``forward" current) and returning along the bell-shaped path on the outside (the ``return" current).\citep{nll06} Fig.~\ref{field} shows a snapshot of the configuration of the magnetic field $\vec{B}$, which indicates that 
a tightly wound central helix is overlapped with the ``forward" current, and a loosely wound helix is overlapped with the ``return" current. Given a helical magnetic field, this axial current-carrying cylindrical plasma column is subject to a current-driven instability (CDI).\citep{hb03,hb05} However, we do not see any visible evidence of any current-driven instability in this case ($\alpha=\sqrt{10}$). The well-known Kruskal-Shafranov criterion \citep{sv57,kjgg58} for MHD kink instability in cylindrical geometry can be written as: \citep{hb05} 
\begin{equation}
\label{ks}
q(a)=\frac{4\pi \psi_p}{LI_{z,{\rm total}}}<1\,.
\end{equation}
where $q$ is the safety factor, $\psi_{p}\approx \pi a^2B_z(a)$ is the total poloidal magnetic flux, $I_{z,\rm {total}}$ is the total axial current, $a$ and $L$ are the column radius and length, and $B_z$ is the axial field component. Safety factors less than $1$ are unstable to the CDI kink mode. The safety factor in this case (Fig.~\ref{jz} and Fig.~\ref{field}) is $q(a)\sim2.8$ at $t=0.5$, which is bigger than $1$. Therefore it is expected to be CDI stable, which is consistent with the simulation results. 
The simulation with $\alpha=15$ gives $q(a)\sim1$ at $t=0.125$, which is marginally unstable to CDI according to Eq.~\ref{ks}. However we do not find evidence of unstable CDI modes in this case either.
``Line tying" (important in the experiment, not present in the simulations reported here due to the ``outflowing" boundary conditions used in this paper) \citep{hp79,eh83,fir06} and other stabilization effects such as ``dynamic relaxation", \citep{nll07}  internal strong rotation \citep{tmt01,nm04} and external gas pressure, \emph{etc.}, could raise the stability threshold. A more detailed stability analysis of the magnetic bubble is beyond the scope of this paper.

\subsection{Angular momentum transport}\label{transport}
Since the bubble is rotating about the $z-$axis initially, the bubble has initial net angular momentum. Conservation of azimuthal angular momentum will slow the bubble's rotation since some angular momentum will be transported to the background plasma. \citep{mku01} It is a key nonlinear plasma physics question to address how this angular momentum evolves. 

For an ideal MHD flow, the azimuthal angular momentum conservation equation in cylindrical coordinates $(r,\varphi,z)$
is: \citep{bh98}
\begin{equation}
\frac{\partial}{\partial t}(\rho r_c v_{\varphi})+\nabla\cdot r_c[\underbrace{\rho v_{\varphi} \vec{v}}_1\underbrace{-B_{\varphi}\vec{B}_p}_2+\underbrace{(p+\frac{B_p^2}{2})\hat{e}_{\varphi}}_3]=0.
\end{equation}
where $\hat{e}_{\varphi}$ is the unit vector in the azimuthal direction, the $p$ subscript refers to a poloidal magnetic-field component (\emph{i.e.}, the $r$ or $z$ component), and $B_p^2=B_r^2+B_z^2$. There are no source terms in this equation, {\em i.e.}, angular momentum may be redistributed in the fluid but never destroyed. The numerical diffusion present in the simulations would transport some angular momentum as well. However, the influence of this transport would be highly limited in the shock/wavefront regions and negligible elsewhere. The first term in the bracket $r_c\rho v_{\varphi}\vec{v}$, the so-called ``advection angular momentum flux" $\Gamma_{\rm advection}$, is the angular momentum flux vector due to the advection, which is defined, in Cartesian coordinates, as:
\begin{equation}
\label{eq:advection}
\vec{\Gamma}_{{\rm advection}} = \rho(xv_y-yv_x)\vec{v}=\rho(xv_y-yv_x)(v_x\hat{x}+v_y\hat{y}+v_z\hat{z})\,,
\end{equation}
where $\hat{x}$, $\hat{y}$ and $\hat{z}$ are the unit vectors in the $x-$, $y-$ and $z-$directions respectively. 
The second term in the bracket $-r_c B_{\varphi}\vec{B}_{p}$, the so-called ``Maxwell angular momentum flux" $\Gamma_{\rm Maxwell}$, is the angular momentum flux vector due to the Lorentz force, which is defined in Cartesian coordinates
as:
\begin{eqnarray}
\label{eq:maxwell}
\vec{\Gamma}_{{\rm Maxell}}& = &-(xB_y-yB_x)\vec{B}_p=-(xB_y-yB_x)(B_r\hat{e}_r+B_z\hat{e}_z)\nonumber\\
&=&-(xB_y-yB_x)[(B_x\cos^2\theta\nonumber\\
&+&B_y\sin\theta\cos\theta)\hat{x}+(B_y\sin^2\theta+B_x\cos\theta\sin\theta)\hat{y}+B_z\hat{z}]\,,
\end{eqnarray}
where $\theta$ is the polar angle with $\tan\theta=x/y$ and $\hat{e}_r$, $\hat{e}_z$ are the radial and axial unit vectors in cylindrical coordinates, respectively.
They both contribute to the angular momentum transport in every direction.
The third term, the so-called ``pressure angular momentum flux" $\Gamma_{\rm pressure}$, is the angular momentum flux vector due to the effective pressure, which is defined in Cartesian coordinates as: 
\[
\vec{\Gamma}_{\rm pressure}=(x\hat{y}-y\hat{x})[p+(B_r^2+B_z^2)/2]\,,
\]
where $B_r=B_x \cos\theta+B_y \sin\theta$. This term does not have a $z-$component, \emph{i.e.}, it only distributes  azimuthal angular momentum in the toroidal plane ($x$-$y$ plane). The total angular momentum flux $\Gamma_{\rm total}$ is defined as:
\[
\vec{\Gamma}_{\rm total}=\vec{\Gamma}_{\rm advection}+\vec{\Gamma}_{\rm Maxwell}+\vec{\Gamma}_{\rm pressure}\,.
\]

\subsubsection{Angular momentum transport in the $x$-$y$ plane} 
\begin{figure}[!htp]
\begin{center}
\subfigure{\scalebox{0.4}{\includegraphics{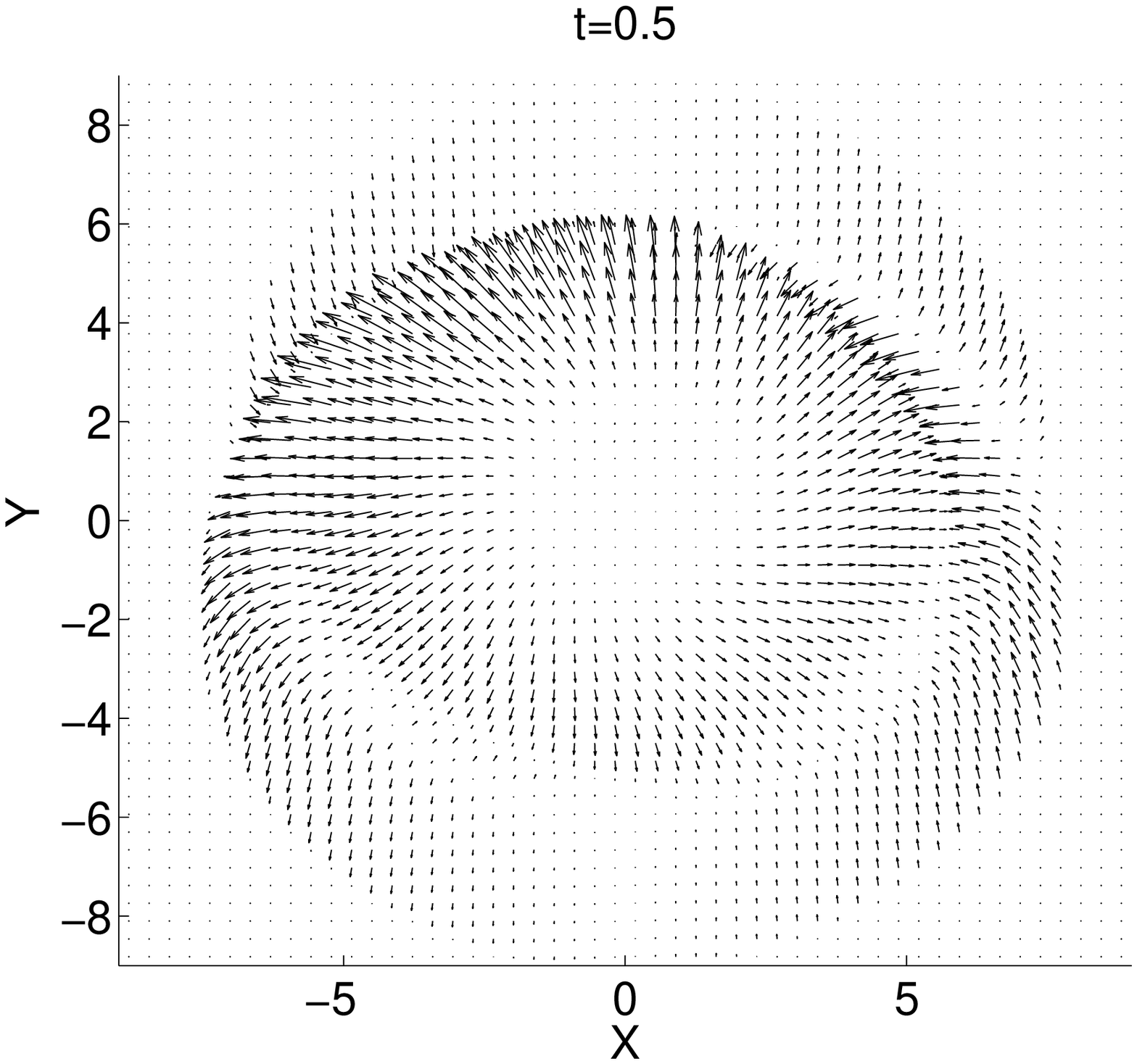}}}$\,$
\subfigure{\scalebox{0.4}{\includegraphics{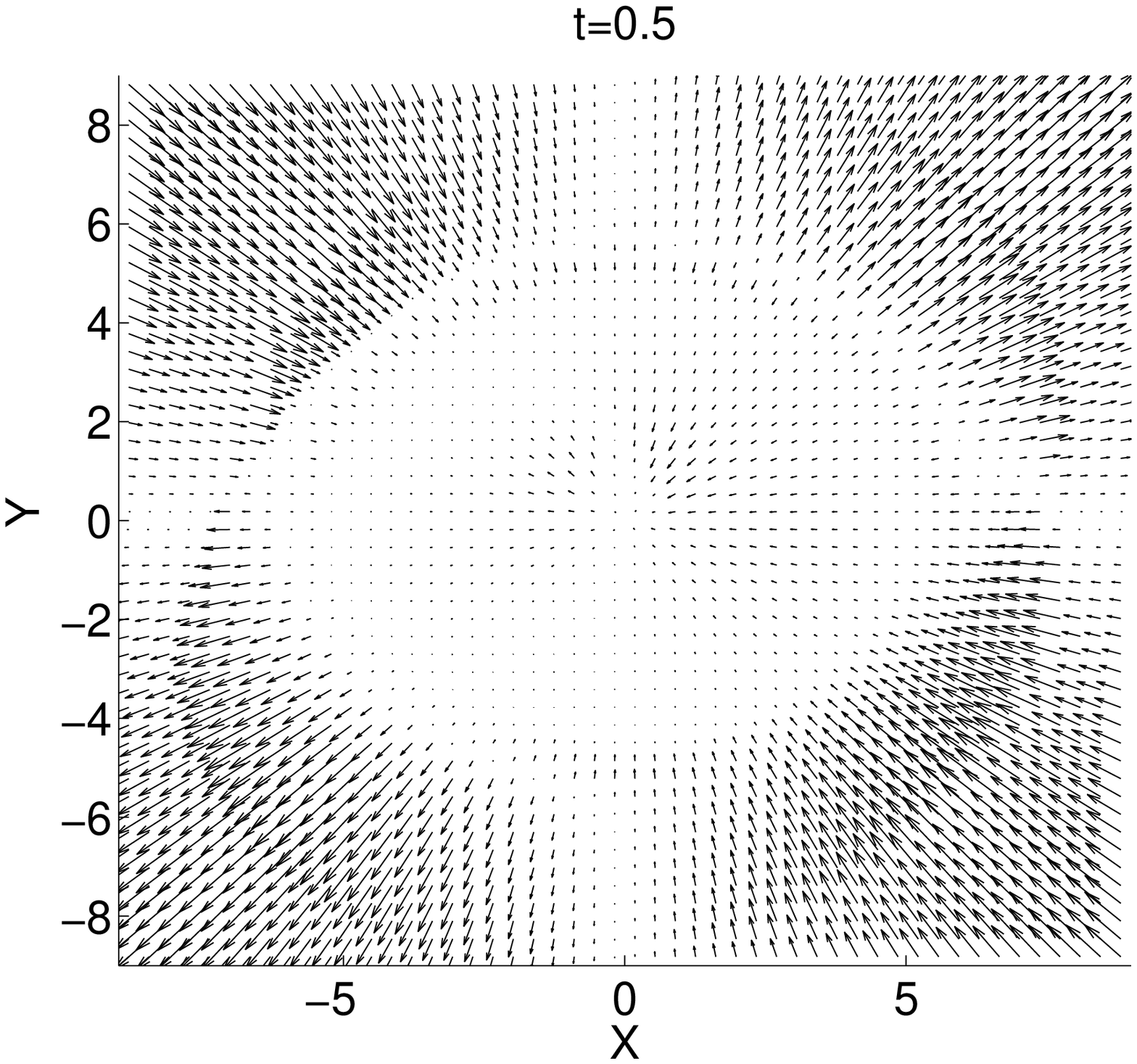}}}\\
\subfigure{\scalebox{0.4}{\includegraphics{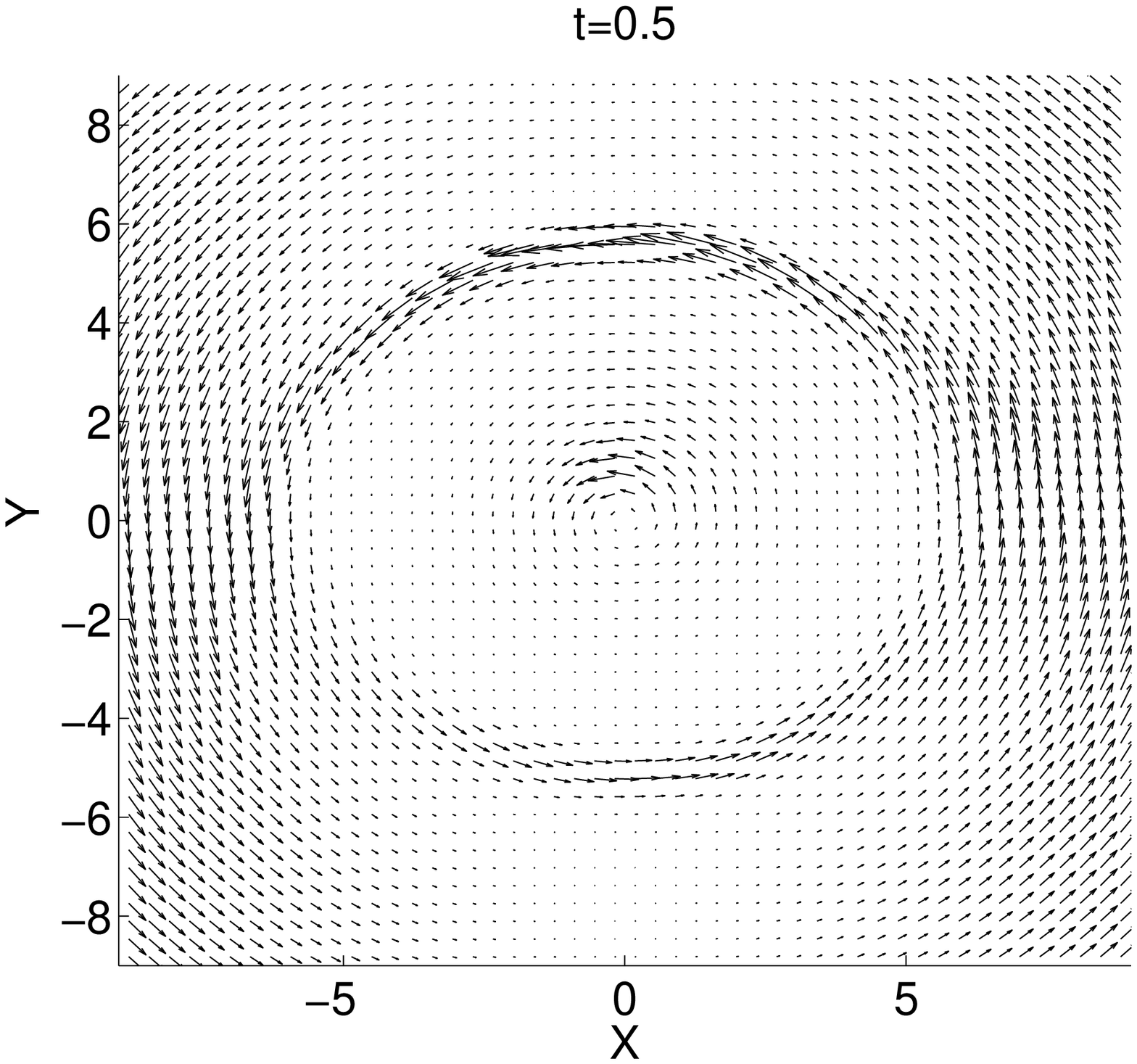}}}$\,$
\subfigure{\scalebox{0.4}{\includegraphics{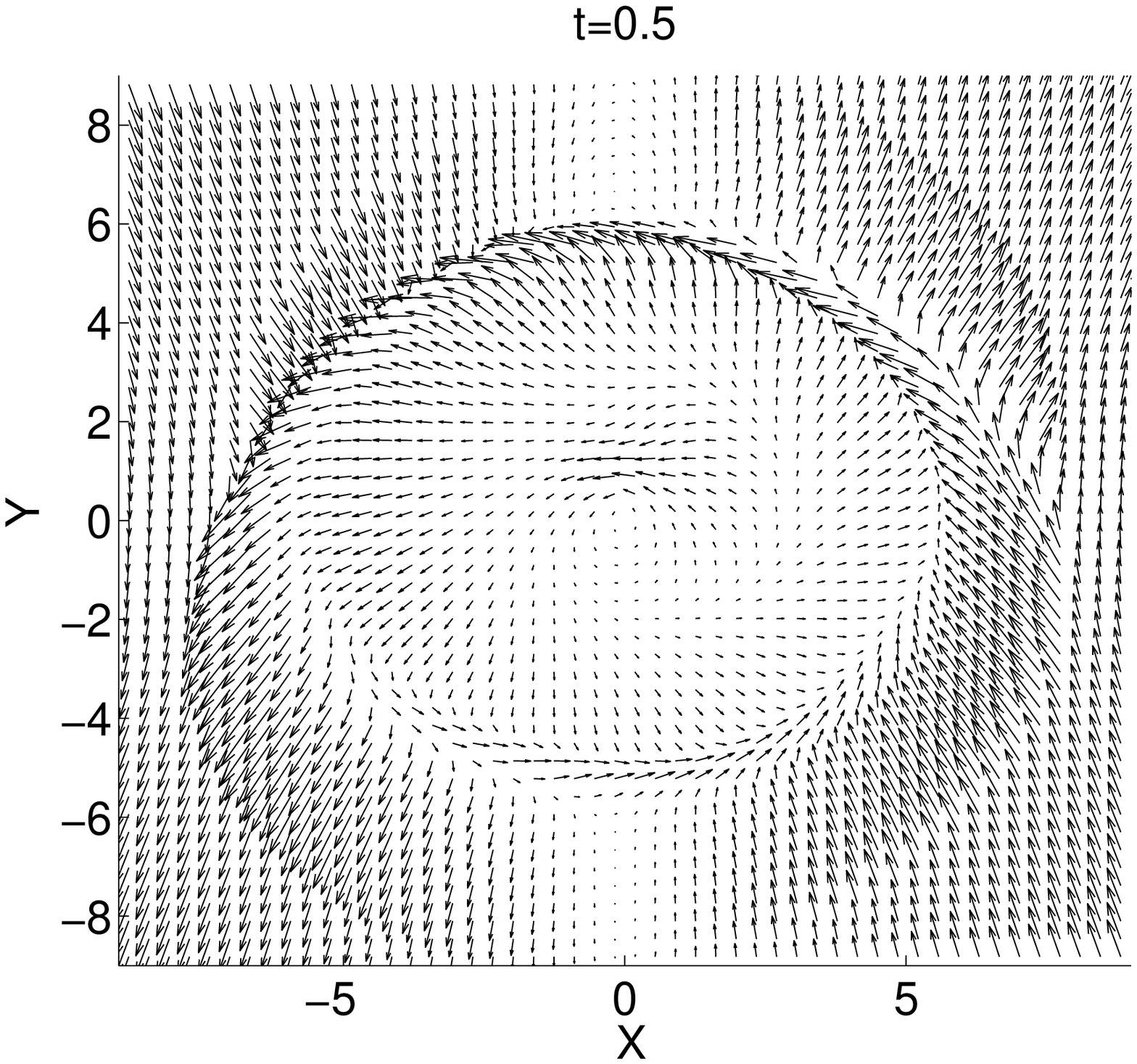}}}\\
\end{center}
\caption{~Vector plot of the angular momentum fluxes due to advection $\Gamma_{\rm advection}$ (top left), Lorentz force $\Gamma_{\rm Maxell}$ (top right), pressure $\Gamma_{\rm pressure}$ (bottom left) and $\Gamma_{\rm Total}$ (bottom right) in the $x$-$y$ plane at $z=-6$ and $t=0.5$.  \label{momentum2} }
\end{figure}

\begin{figure}[!htp]
\begin{center}
\subfigure{\scalebox{0.4}{\includegraphics{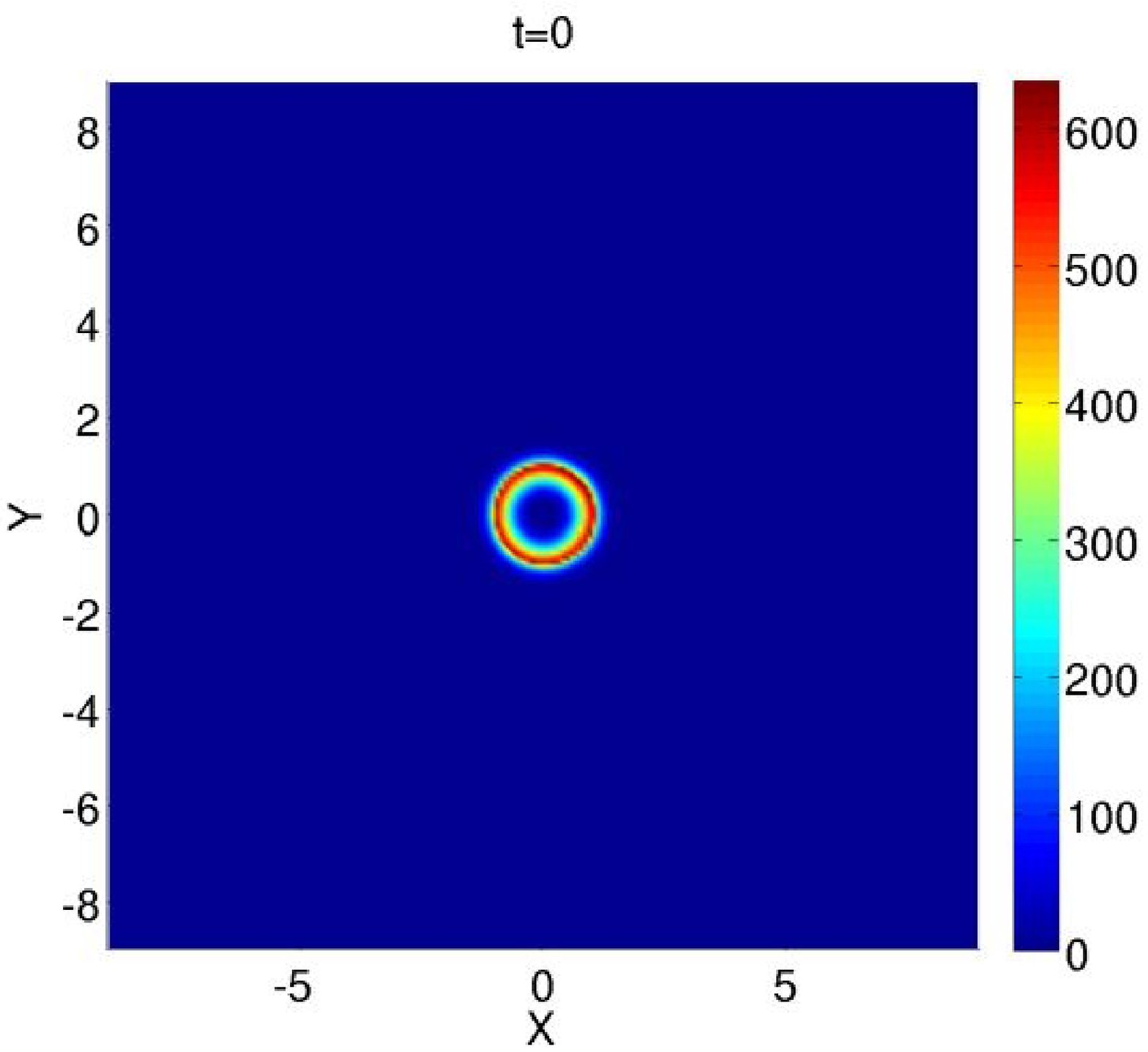}}}$\,$
\subfigure{\scalebox{0.4}{\includegraphics{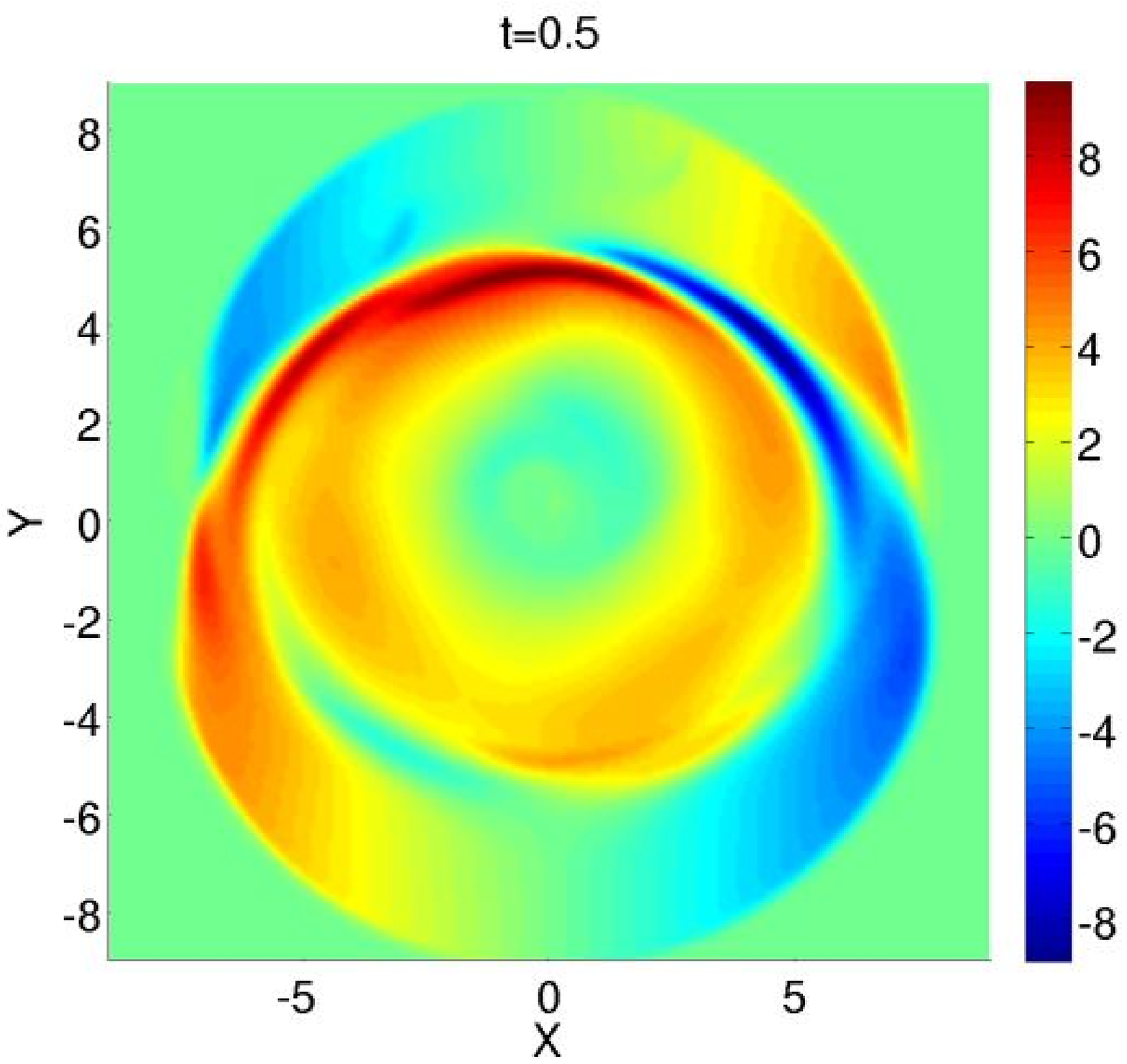}}}
\end{center}
\caption{(color) Specific azimuthal angular momentum $\rho r_c v_{\varphi}$ in the $x$-$y$ plane at two times. Left: $t=0$, $z=-7.5$.  Right: $t=0.5$, $z=-6$.  \label{angular2} }
\end{figure}

Figure~\ref{momentum2} displays vector plots of $\Gamma_{\rm advection}$, $\Gamma_{\rm Maxwell}$, $\Gamma_{\rm pressure}$ and $\Gamma_{\rm total}$ in the toroidal plane, \emph{i.e.}, $x-y$ plane at $z=-6$, which
coincides with the bubble.  Pressure simply transports angular momentum in an anti-clockwise direction in this plane. Inside the bubble, advection transports angular momentum outward, which is due to the expansion of the bubble. The Maxwell angular momentum flux only has a radial component in the plane (Eq.~\ref{eq:maxwell}). This flux is dominant outside the bubble since the radial bubble field component decreases so quickly that it is much smaller than the background field at $t=0.5$ [see also Fig.~\ref{field_xy}(right)] due to the quick relaxation (Sec.~\ref{balance}). Interestingly the Lorentz force transports angular momentum inward at the top left and bottom right regions and
outward at the top right and bottom left regions. Therefore the total effect is to transport
angular momentum: (1) angular momentum outward inside the bubble; (2) along the negative $y-$axis for $x<0$ and along the positive $y-$axis for $x>0$ outside the bubble. The maximum angular momentum transport happens at the edge of the bubble. Inside the bubble some angular momentum has been transported from the center of the bubble to the edge of the bubble, while outside some angular momentum has been transported from the top left to the bottom left regions, and from the bottom right to the top right regions [Fig.~\ref{angular2}(right)]. Thus, the uniformly rotating bubble expands and its inner region ceases to rotate and then rotates oppositely in the long run, while the neighboring plasma starts to rotate differentially.  The top right and bottom left regions rotate in the same direction as the original bubble, while the top left and bottom right regions rotate in the opposite direction (Fig.~\ref{angular2}), which results in shears. This explains why, between the shock and wavefront, the advection transports the angular momentum in negatively when $x<0$ while positively when $x>0$ since the shock is always propagating outward (Eq.~\ref{eq:advection}).

\subsubsection{Angular momentum transport in the $x$-$z$ plane}
\begin{figure}[!htp]
\begin{center}
\subfigure{\scalebox{0.4}{\includegraphics{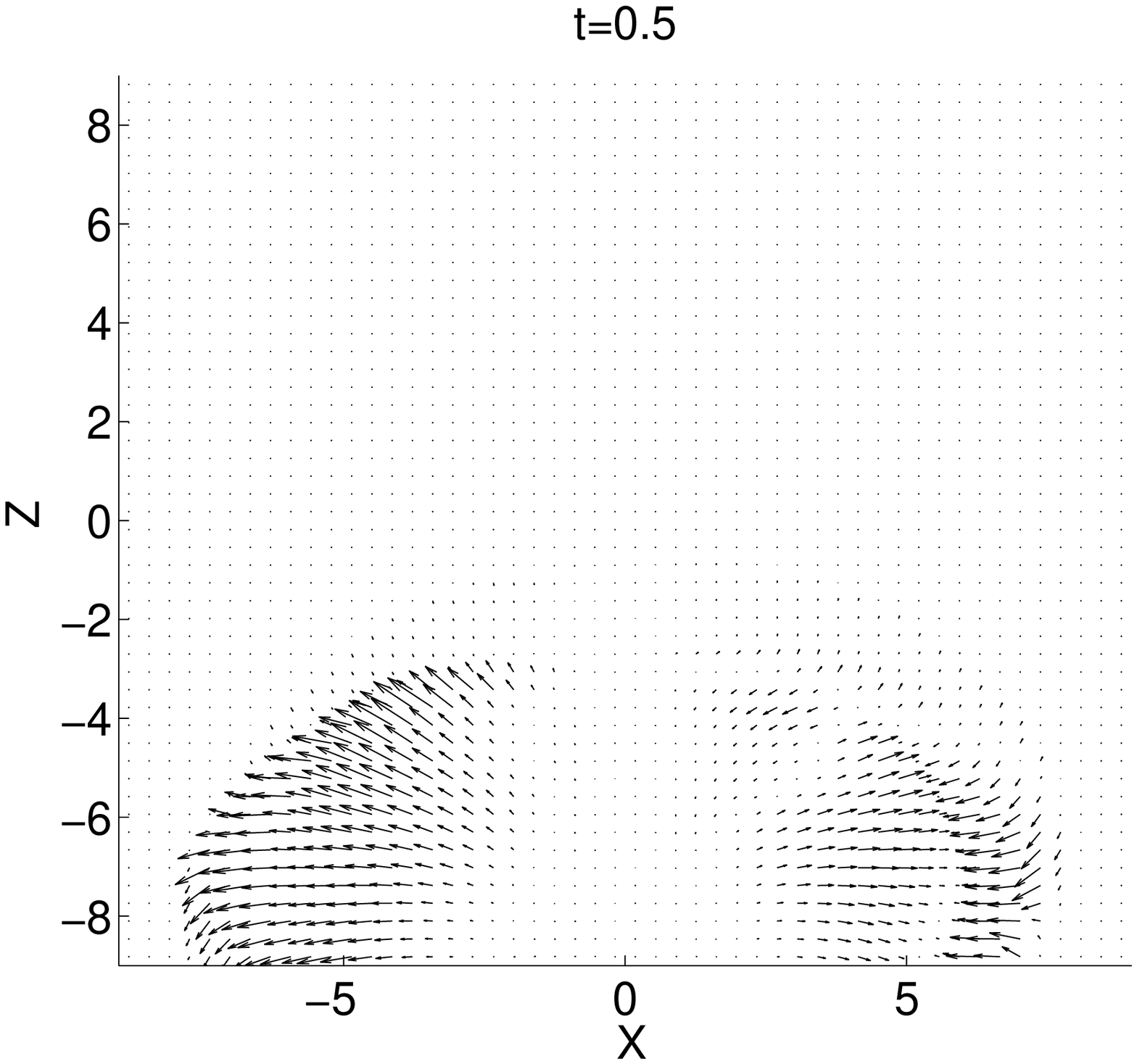}}}$\,$
\subfigure{\scalebox{0.4}{\includegraphics{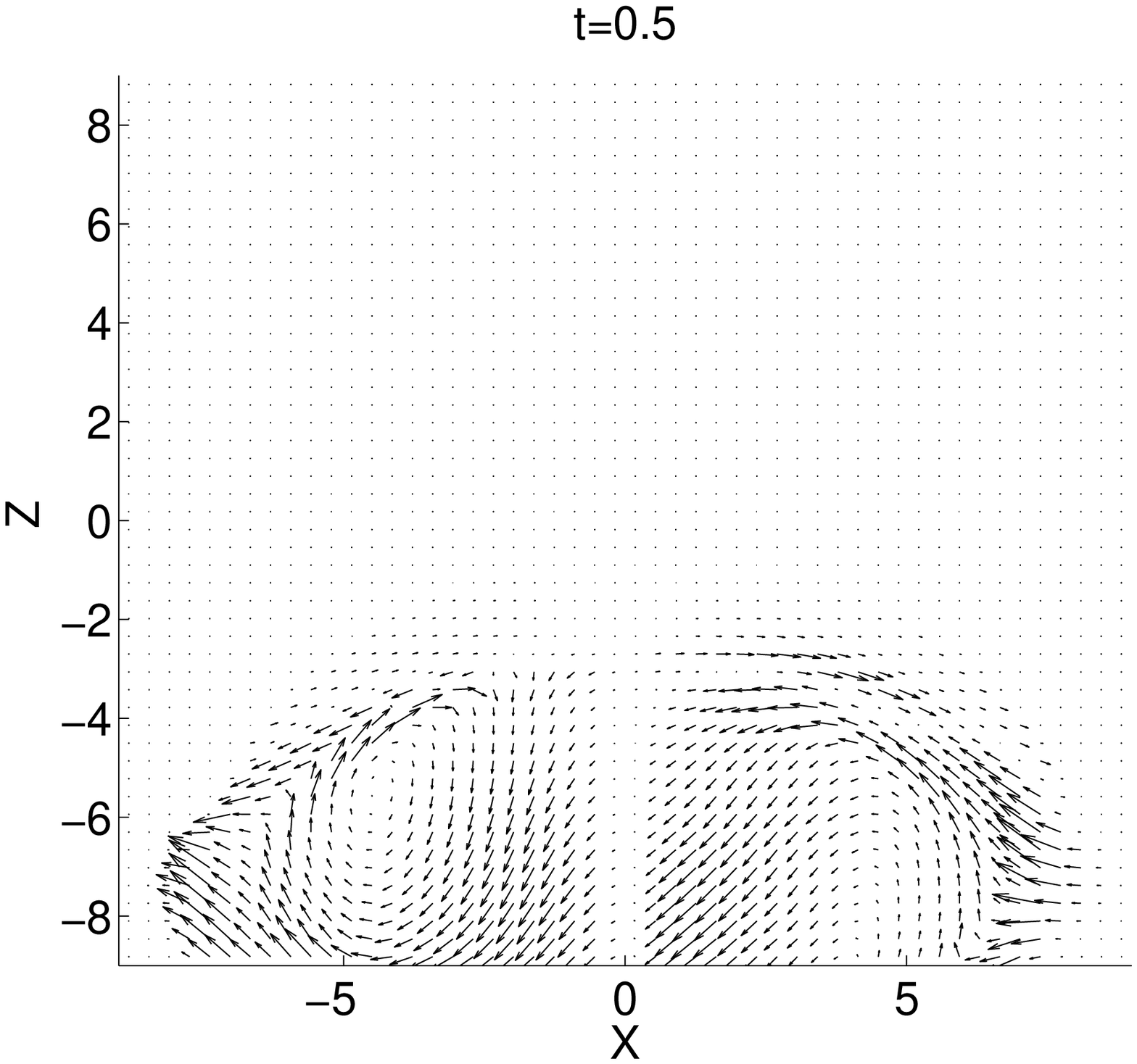}}}\\
\subfigure{\scalebox{0.4}{\includegraphics{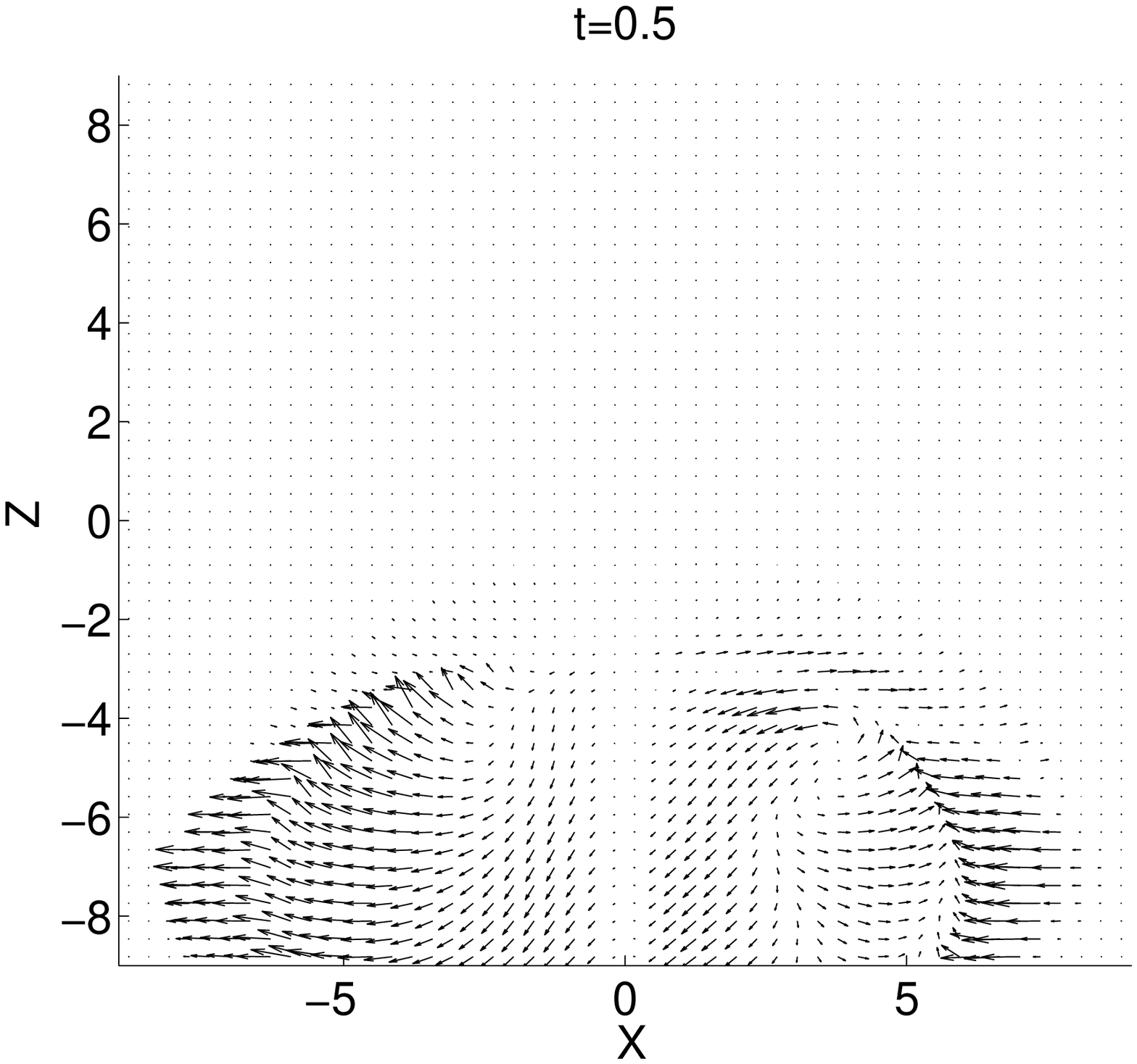}}}
\end{center}
\caption{~Vector plot of the angular momentum fluxes due to advection $\Gamma_{\rm advection}$ (top left), Lorentz force $\Gamma_{\rm Maxell}$ (top right) and $\Gamma_{\rm Total}=\Gamma_{\rm advection}+\Gamma_{\rm Maxwell}$ (bottom) in the $x$-$z$ plane at $y=0$ and $t=0.5$.  \label{momentum} }
\end{figure}

\begin{figure}[!htp]
\begin{center}
\subfigure{\scalebox{0.4}{\includegraphics{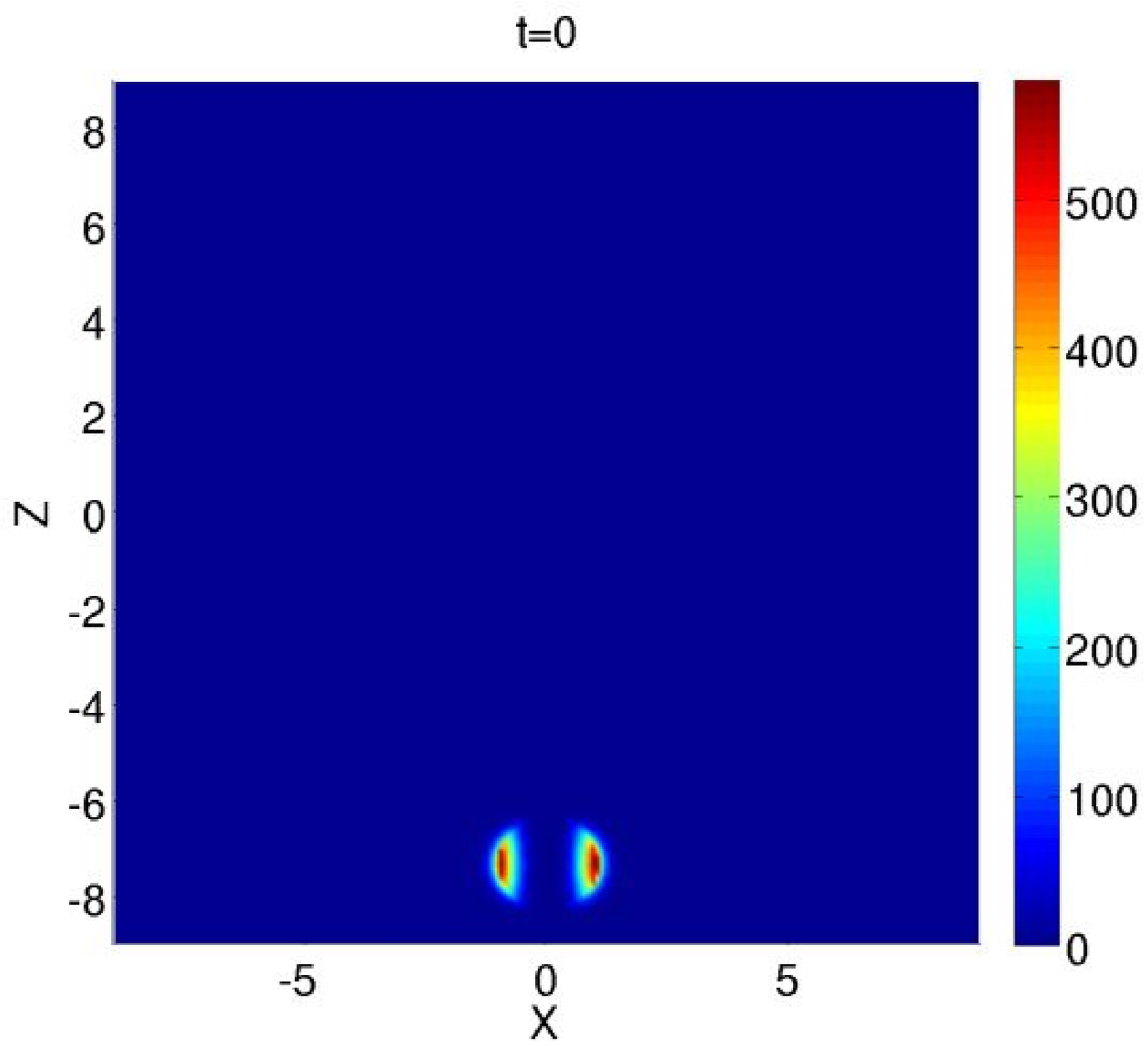}}}$\,$
\subfigure{\scalebox{0.4}{\includegraphics{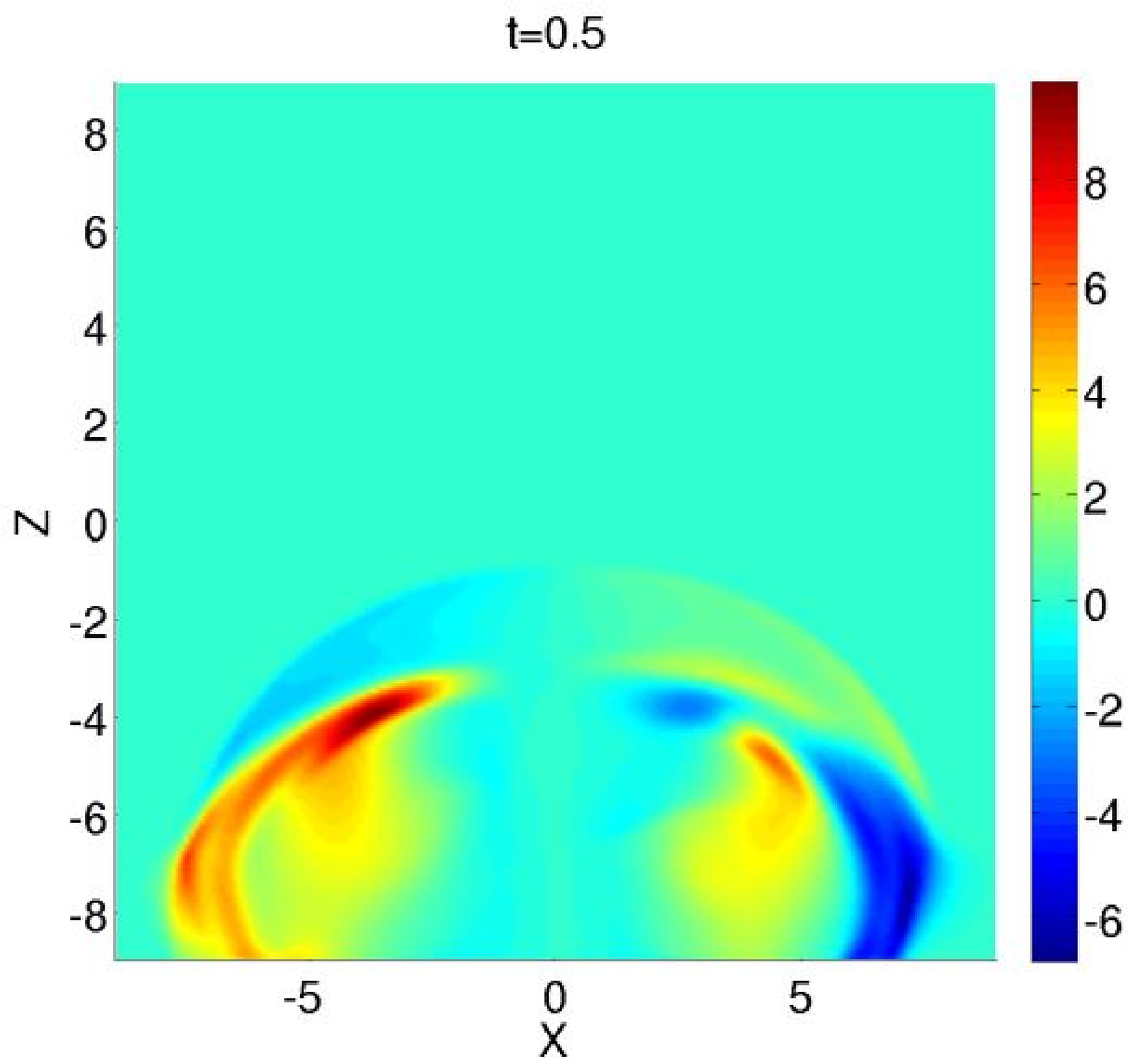}}}
\end{center}
\caption{~(color) Contour plot of the specific azimuthal angular momentum $\rho r_c v_{\varphi}$ in the $x$-$z$ plane at $y=0$ at different times. Left: $t=0$.  Right: $t=0.5$. Net angular momentum is transported from the right hand side to the left hand side of the bubble edge. \label{angular} }
\end{figure}

Angular momentum transport in the poloidal plane, \emph{i.e.}, the $x$-$z$ plane at $y=0$, is presented in Fig.~\ref{momentum} ($\Gamma_{\rm pressure}$ is zero on this plane). Inside the bubble, $\Gamma_{\rm advection}$ due to the expansion of the bubble transports angular momentum outward normal to the wavefront, while $\Gamma_{\rm Maxwell}$ due to the bubble field redistributes angular momentum inside, transporting angular momentum clockwise on the left hand side and anti-closewise on the right hand side. This can be understood from the spheromak-like magnetic field configuration of the bubble. 
The total effect is to transport net angular momentum from the right hand side to the left hand side of the bubble edge, leading to positive angular momentum on the left hand side and negative angular momentum on the right hand side of the bubble edge (Fig.~\ref{angular}).  


\section{Summary \& discussions}\label{discussions} 
In this paper we presented initial nonlinear ideal MHD simulation results of the expansion of a magnetic bubble into a lower pressure weakly magnetized background plasma. The simulations mimic the ongoing experiment PBEX, except that we use simplified "out-flowing" boundary conditions and ignore collisional effects. A high-density magnetized bubble is
injected into a cylindrical background plasma. The bubble evolution is dependent on $\alpha$, with larger $\alpha$ resulting in an axially expanding bubble like a growing ``mushroom" and smaller $\alpha$ producing a ``crab-like"
shape expanding normal to the direction of propagation. The expansion of the bubble generates one leading MHD shock and one trailing reverse slow-mode compressible MHD wavefront. The shock/wavefront speed is independent of the injection velocity for injection velocities similar or below $V_A$. In the $x$-$z$ plane anti-parallel reconnection takes place on the left hand side of the wavefront, where the bubble field is opposite to the background field, while in the $x$-$y$ plane the reconnection takes place if $y>0$. The azimuthal angular momentum is transported outward from the center to the edge of the bubble by advection, and from the right hand side to the left hand side of the bubble edge by the Maxwell torque.  Outside the bubble, angular momentum is transported from the top left to the bottom left and from the bottom right to top right by the combination of Maxwell and pressure angular momentum fluxes. The initial uniformly rotating bubble quickly evolves into a quasi-force-free state, and the center of the bubble ceases to rotate and starts to rotate oppositely in the long run, while the outside neighboring plasmas start to rotate differentially:  the top right and bottom left regions possessing the same rotating direction as the original bubbe while the top left and bottom right regions possessing the opposite rotating direction, which forms shears in the system. 

The background magnetic field breaks the symmetry of the system along the propagation axis. For comparison,
simulations with 2 orders of magnitude lower background field (and correspondingly lower density and temperature in order to preserve the equilibrium of the background plasmas) were also performed. The results show much better symmetry along the propagation axis.  More detailed studies of the role of the background field will be the subject
of future work.

From Table~\ref{toroidal}, the resistive dissipation time due to numerical diffusion is inferred to be $\tau_{\rm res}\sim 1.2$, which is longer than the time ($t=0.5$) at which the shock has reached the $x$ and $y$ boundaries. The Alfv\'en time can be calculated as $\tau_{a}=L_{\rm res}/V_{\rm A, res}$, where $L_{\rm res}$ ($\sim0.5$) is the typical length of the reconnection layer and $V_{\rm A, res}$ is the Alfv\'en speed ($\sim8$) at the reconnection layer. This gives $\tau_{a}\sim0.06$. Thus the effective Lundquist number $S_{\rm effective}$ in the simulations is around $S_{\rm effective}=\tau_{\rm res}/\tau_{a}\sim20$. Please note that this is the Lundquist number associated with the reconnection layer and the numerical diffusion used here is the upper limit of the numerical diffusion in the simulations. The estimate of the mean experimental Lundquist number $S_{\rm experiment}=\left<V_{A}^0L_0/\eta_{\rm plasma}\right>$ is around $200$, where $L_0\sim18$ is the characteristic length of the experimental facility, $V_{A}^{0}$ is the initial Alfv\'en speed, $\eta_{\rm plasma}$ is the magnetic resistivity of the plasma at the initial state based on Braginskii's formula and $\left<...\right>$ indicates the volume average. Although there is one order of magnitude difference between them, we expect our ideal simulations with numerical diffusion to give a reasonable estimate of the physical quantities in the real experiment.

\begin{figure}[!htp]
\begin{center}
\subfigure{\scalebox{0.4}{\includegraphics{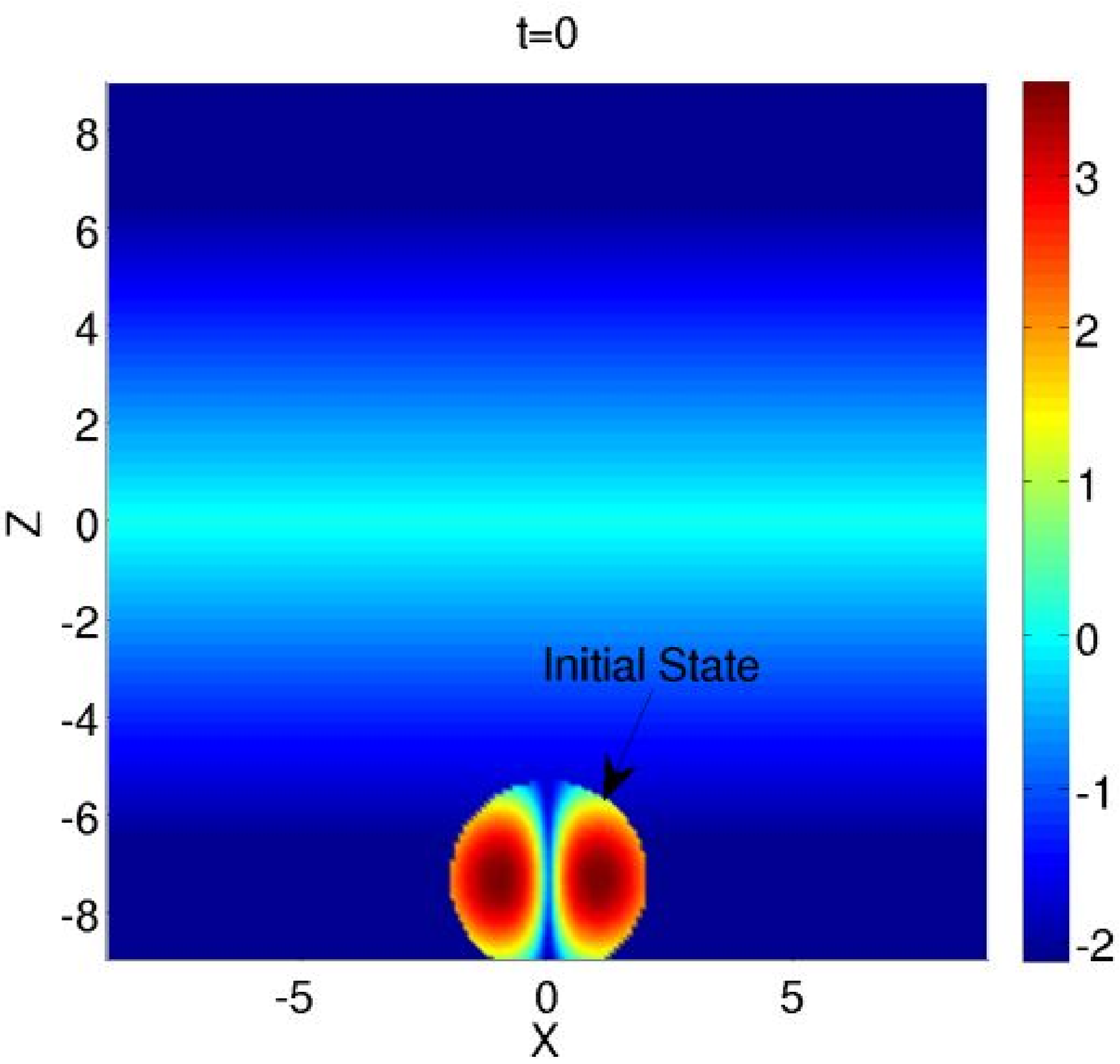}}}$\,$
\subfigure{\scalebox{0.4}{\includegraphics{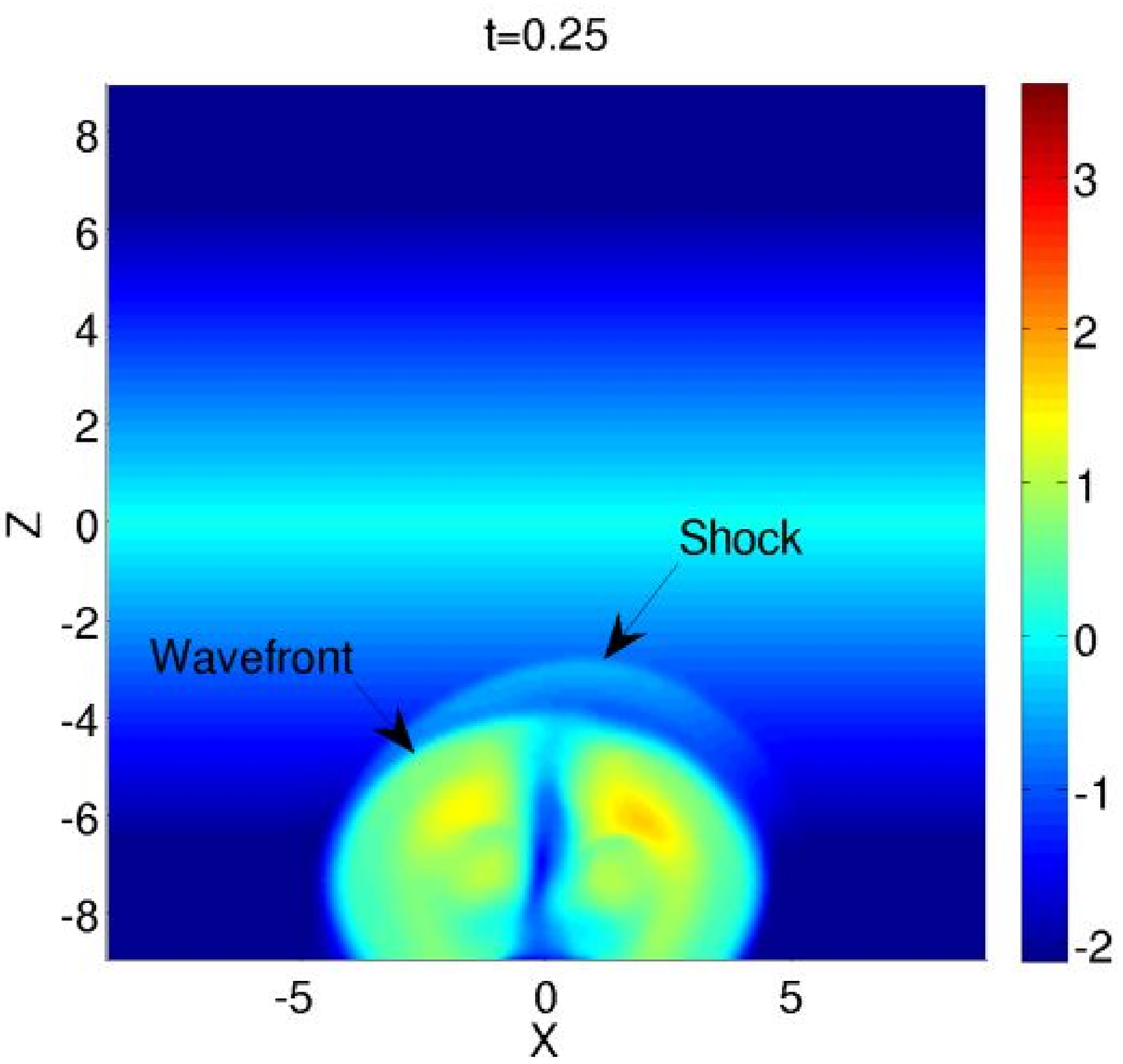}}}\\
\subfigure{\scalebox{0.4}{\includegraphics{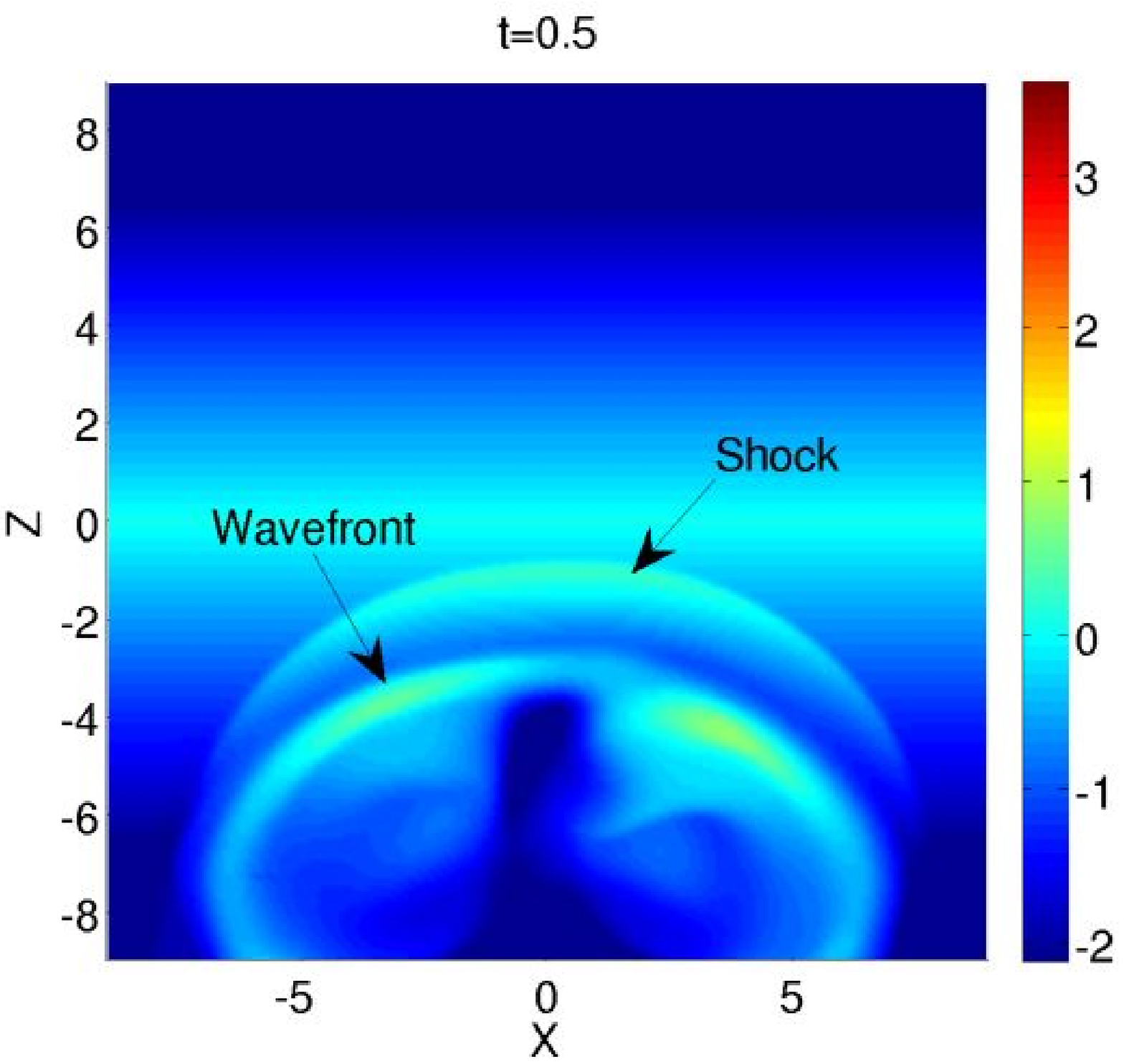}}}$\,$
\subfigure{\scalebox{0.4}{\includegraphics{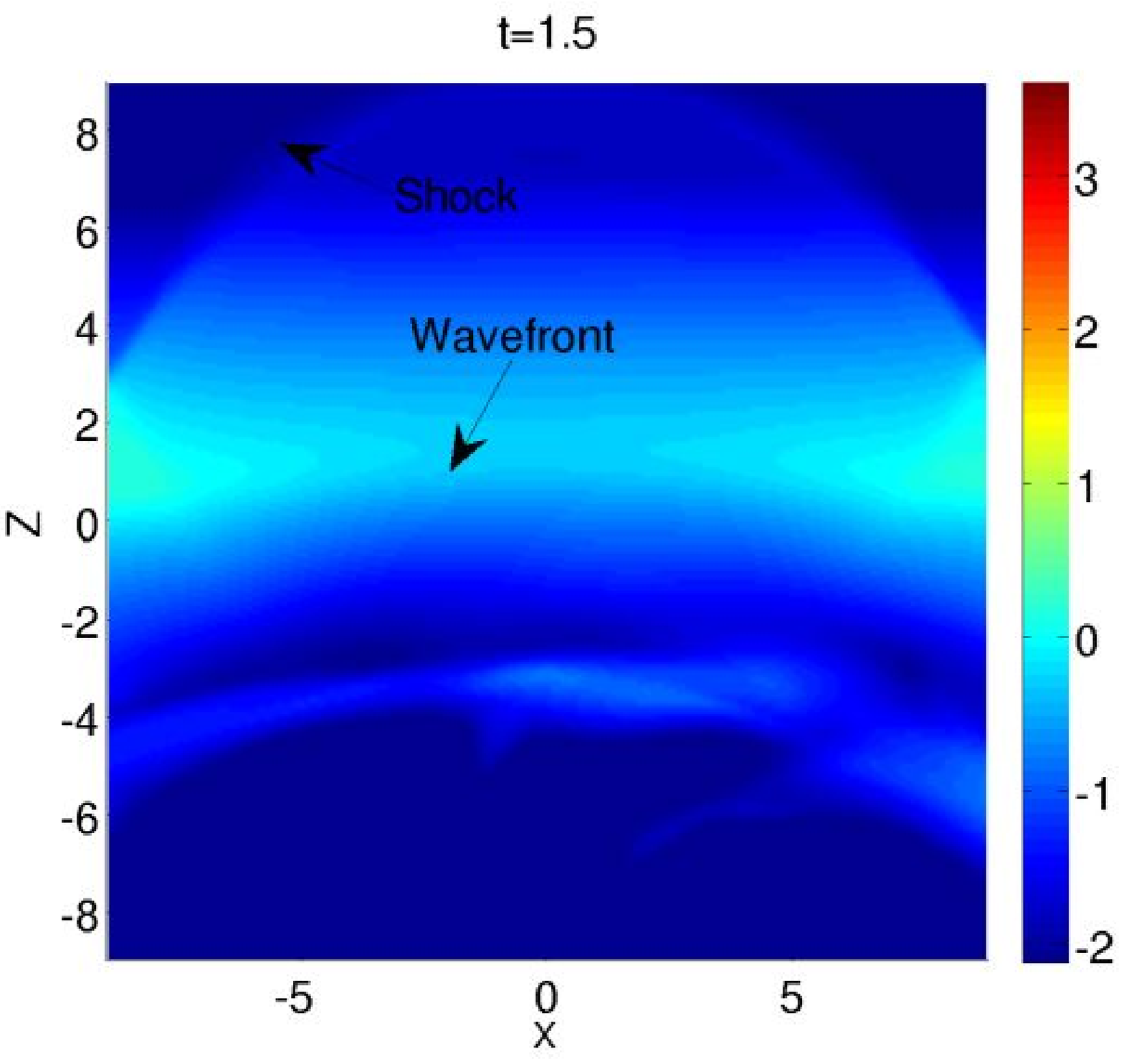}}}
\end{center}
\caption{~(color) Density (natural logarithmic scale) in the $x$-$z$ plane as a function of time ($\alpha=\sqrt{10}$ and
$V_{\rm inj}=0.18V_{A,0}$). \label{overview} }
\end{figure}

Another issue important in the experiment is the transit time. Although the value of the transit time is somewhat related to the boundary conditions, our simulations with simplified boundary condition show that the background plasma column radius $(r_p=20\unit{cm})$ is large enough to allow the bubble to relax substantially.
From Fig.~\ref{overview}, it is seen that the bubble is still inside the background plasma at $t=1.5$.
However, better boundary conditions are needed for the simulations to be meaningful after the shock has reach the boundaries.

The appearance of the MHD shock and wavefront suggests that our experimental facility may provide a unique opportunity to study MHD shocks in a laboratory plasma.  
However, we emphasize that these conclusions are based on ideal simulations (with numerical diffusion) and that the boundary conditions are not realistic. This paper 
is intended as a preliminary exploration of PBEX\@. We have not attempted to model many of the 
complexities of a realistic experiment. In future papers, we will study collisional effects and boundary conditions closer to those of the planned experiment; work in progress indicates that these will modify the results.

\acknowledgments{This work was supported by the Los Alamos Directed Research and Development (LDRD) Program under
Department of Energy contract No.\ DE-AC52-06NA25396.}


\begin{thebibliography}{32}
\expandafter\ifx\csname natexlab\endcsname\relax\def\natexlab#1{#1}\fi
\expandafter\ifx\csname bibnamefont\endcsname\relax
  \def\bibnamefont#1{#1}\fi
\expandafter\ifx\csname bibfnamefont\endcsname\relax
  \def\bibfnamefont#1{#1}\fi
\expandafter\ifx\csname citenamefont\endcsname\relax
  \def\citenamefont#1{#1}\fi
\expandafter\ifx\csname url\endcsname\relax
  \def\url#1{\texttt{#1}}\fi
\expandafter\ifx\csname urlprefix\endcsname\relax\def\urlprefix{URL }\fi
\providecommand{\bibinfo}[2]{#2}
\providecommand{\eprint}[2][]{\url{#2}}

\bibitem[{\citenamefont{B\^izan et~al.}(2004)\citenamefont{B\^izan, Rafferty,
  and McNamara}}]{brm04}
\bibinfo{author}{\bibfnamefont{L.}~\bibnamefont{B\^izan}},
  \bibinfo{author}{\bibfnamefont{D.~A.} \bibnamefont{Rafferty}},
  \bibnamefont{and} \bibinfo{author}{\bibfnamefont{B.~R.}
  \bibnamefont{McNamara}}, \bibinfo{journal}{Astrophys. J.}
  \textbf{\bibinfo{volume}{607}}, \bibinfo{pages}{800} (\bibinfo{year}{2004}).

\bibitem[{\citenamefont{McNamara et~al.}(2000)\citenamefont{McNamara, Wise,
  Nulsen, David, Sarazin, Bautz, Markevitch, Vikhlinin, Forman, Jones
  et~al.}}]{mwn00}
\bibinfo{author}{\bibfnamefont{B.~R.} \bibnamefont{McNamara}},
  \bibinfo{author}{\bibfnamefont{M.}~\bibnamefont{Wise}},
  \bibinfo{author}{\bibfnamefont{P.~E.~J.} \bibnamefont{Nulsen}},
  \bibinfo{author}{\bibfnamefont{L.~P.} \bibnamefont{David}},
  \bibinfo{author}{\bibfnamefont{C.~L.} \bibnamefont{Sarazin}},
  \bibinfo{author}{\bibfnamefont{M.}~\bibnamefont{Bautz}},
  \bibinfo{author}{\bibfnamefont{M.}~\bibnamefont{Markevitch}},
  \bibinfo{author}{\bibfnamefont{A.}~\bibnamefont{Vikhlinin}},
  \bibinfo{author}{\bibfnamefont{W.~R.} \bibnamefont{Forman}},
  \bibinfo{author}{\bibfnamefont{C.}~\bibnamefont{Jones}},
  \bibnamefont{et~al.}, \bibinfo{journal}{Astrophys. J.}
  \textbf{\bibinfo{volume}{534}}, \bibinfo{pages}{L135} (\bibinfo{year}{2000}).

\bibitem[{\citenamefont{McNamara et~al.}(2005)\citenamefont{McNamara, Nulsen,
  Wise, Rafferty, Carilli, Sarazin, and Blanton}}]{mnw05}
\bibinfo{author}{\bibfnamefont{B.~R.} \bibnamefont{McNamara}},
  \bibinfo{author}{\bibfnamefont{P.~E.~J.} \bibnamefont{Nulsen}},
  \bibinfo{author}{\bibfnamefont{M.~W.} \bibnamefont{Wise}},
  \bibinfo{author}{\bibfnamefont{D.~A.} \bibnamefont{Rafferty}},
  \bibinfo{author}{\bibfnamefont{C.}~\bibnamefont{Carilli}},
  \bibinfo{author}{\bibfnamefont{C.~L.} \bibnamefont{Sarazin}},
  \bibnamefont{and} \bibinfo{author}{\bibfnamefont{E.~L.}
  \bibnamefont{Blanton}}, \bibinfo{journal}{Nature}
  \textbf{\bibinfo{volume}{433}}, \bibinfo{pages}{45} (\bibinfo{year}{2005}).

\bibitem[{\citenamefont{Norman et~al.}(1988)\citenamefont{Norman, Burns, and
  Sulkanen}}]{nbs88}
\bibinfo{author}{\bibfnamefont{M.~L.} \bibnamefont{Norman}},
  \bibinfo{author}{\bibfnamefont{J.~O.} \bibnamefont{Burns}}, \bibnamefont{and}
  \bibinfo{author}{\bibfnamefont{M.~E.} \bibnamefont{Sulkanen}},
  \bibinfo{journal}{Nature} \textbf{\bibinfo{volume}{335}},
  \bibinfo{pages}{146} (\bibinfo{year}{1988}).

\bibitem[{\citenamefont{Marti et~al.}(1994)\citenamefont{Marti, Mueller, and
  Ibanez}}]{mmi94}
\bibinfo{author}{\bibfnamefont{J.~M.} \bibnamefont{Marti}},
  \bibinfo{author}{\bibfnamefont{E.}~\bibnamefont{Mueller}}, \bibnamefont{and}
  \bibinfo{author}{\bibfnamefont{J.~M.} \bibnamefont{Ibanez}},
  \bibinfo{journal}{A\&A} \textbf{\bibinfo{volume}{281}}, \bibinfo{pages}{L9}
  (\bibinfo{year}{1994}).

\bibitem[{\citenamefont{Clarke}(1996)}]{cd96}
\bibinfo{author}{\bibfnamefont{D.~A.} \bibnamefont{Clarke}}, in
  \emph{\bibinfo{booktitle}{ASP Conf. Ser. 100.}}, edited by
  \bibinfo{editor}{\bibfnamefont{P.}~\bibnamefont{Hardee}},
  \bibinfo{editor}{\bibfnamefont{B.}~\bibnamefont{A.H.}}, \bibnamefont{and}
  \bibinfo{editor}{\bibfnamefont{J.}~\bibnamefont{Zensus}}
  (\bibinfo{address}{San Francisco: ASP}, \bibinfo{year}{1996}), p.
  \bibinfo{pages}{311}.

\bibitem[{\citenamefont{Bodo et~al.}(1998)\citenamefont{Bodo, Rossi, Massaglia,
  Ferrari, Malagoli, and Rosner}}]{brm98}
\bibinfo{author}{\bibfnamefont{G.}~\bibnamefont{Bodo}},
  \bibinfo{author}{\bibfnamefont{P.}~\bibnamefont{Rossi}},
  \bibinfo{author}{\bibfnamefont{S.}~\bibnamefont{Massaglia}},
  \bibinfo{author}{\bibfnamefont{A.}~\bibnamefont{Ferrari}},
  \bibinfo{author}{\bibfnamefont{A.}~\bibnamefont{Malagoli}}, \bibnamefont{and}
  \bibinfo{author}{\bibfnamefont{R.}~\bibnamefont{Rosner}},
  \bibinfo{journal}{A\&A} \textbf{\bibinfo{volume}{333}}, \bibinfo{pages}{1117}
  (\bibinfo{year}{1998}).

\bibitem[{\citenamefont{Tregillis et~al.}(2004)\citenamefont{Tregillis, Jones,
  and Ryu}}]{tjr04}
\bibinfo{author}{\bibfnamefont{I.~L.} \bibnamefont{Tregillis}},
  \bibinfo{author}{\bibfnamefont{T.~W.} \bibnamefont{Jones}}, \bibnamefont{and}
  \bibinfo{author}{\bibfnamefont{D.}~\bibnamefont{Ryu}}, \bibinfo{journal}{ApJ}
  \textbf{\bibinfo{volume}{601}}, \bibinfo{pages}{778} (\bibinfo{year}{2004}).

\bibitem[{\citenamefont{Owen et~al.}(2000)\citenamefont{Owen, Eilek, and
  Kassim}}]{oek00}
\bibinfo{author}{\bibfnamefont{F.~N.} \bibnamefont{Owen}},
  \bibinfo{author}{\bibfnamefont{J.~A.} \bibnamefont{Eilek}}, \bibnamefont{and}
  \bibinfo{author}{\bibfnamefont{N.~E.} \bibnamefont{Kassim}},
  \bibinfo{journal}{Astrophys. J.} \textbf{\bibinfo{volume}{543}},
  \bibinfo{pages}{611} (\bibinfo{year}{2000}).

\bibitem[{\citenamefont{Kronberg et~al.}(2001)\citenamefont{Kronberg, Dufton,
  Li, and Colgate}}]{kdl01}
\bibinfo{author}{\bibfnamefont{P.~P.} \bibnamefont{Kronberg}},
  \bibinfo{author}{\bibfnamefont{Q.~W.} \bibnamefont{Dufton}},
  \bibinfo{author}{\bibfnamefont{H.}~\bibnamefont{Li}}, \bibnamefont{and}
  \bibinfo{author}{\bibfnamefont{S.~A.} \bibnamefont{Colgate}},
  \bibinfo{journal}{Astrophys, J.} \textbf{\bibinfo{volume}{560}},
  \bibinfo{pages}{178} (\bibinfo{year}{2001}).

\bibitem[{\citenamefont{Furlanetto and Loeb}(2001)}]{fl01}
\bibinfo{author}{\bibfnamefont{S.~R.} \bibnamefont{Furlanetto}}
  \bibnamefont{and} \bibinfo{author}{\bibfnamefont{A.}~\bibnamefont{Loeb}},
  \bibinfo{journal}{Astrophys. J.} \textbf{\bibinfo{volume}{556}},
  \bibinfo{pages}{619} (\bibinfo{year}{2001}).

\bibitem[{\citenamefont{Croston et~al.}(2005)\citenamefont{Croston, Hardcastle,
  Harris, Belsole, Birkinshaw, and Worrall}}]{chh05}
\bibinfo{author}{\bibfnamefont{J.~H.} \bibnamefont{Croston}},
  \bibinfo{author}{\bibfnamefont{M.~J.} \bibnamefont{Hardcastle}},
  \bibinfo{author}{\bibfnamefont{D.~E.} \bibnamefont{Harris}},
  \bibinfo{author}{\bibfnamefont{E.}~\bibnamefont{Belsole}},
  \bibinfo{author}{\bibfnamefont{M.}~\bibnamefont{Birkinshaw}},
  \bibnamefont{and} \bibinfo{author}{\bibfnamefont{D.~M.}
  \bibnamefont{Worrall}}, \bibinfo{journal}{Astrophys. J.}
  \textbf{\bibinfo{volume}{626}}, \bibinfo{pages}{733} (\bibinfo{year}{2005}).

\bibitem[{\citenamefont{Li et~al.}(2006)\citenamefont{Li, Lapenta, Finn, Li,
  and Colgate}}]{llf06}
\bibinfo{author}{\bibfnamefont{H.}~\bibnamefont{Li}},
  \bibinfo{author}{\bibfnamefont{G.}~\bibnamefont{Lapenta}},
  \bibinfo{author}{\bibfnamefont{J.~M.} \bibnamefont{Finn}},
  \bibinfo{author}{\bibfnamefont{S.}~\bibnamefont{Li}}, \bibnamefont{and}
  \bibinfo{author}{\bibfnamefont{S.~A.} \bibnamefont{Colgate}},
  \bibinfo{journal}{Astrophys. J.} \textbf{\bibinfo{volume}{643}},
  \bibinfo{pages}{92} (\bibinfo{year}{2006}).

\bibitem[{\citenamefont{Tang}(2007)}]{tx07}
\bibinfo{author}{\bibfnamefont{X.~Z.} \bibnamefont{Tang}}
  (\bibinfo{year}{2007}), \bibinfo{note}{submitted to Astrophys. J.}

\bibitem[{\citenamefont{Nakamura and Meier}(2004)}]{nm04}
\bibinfo{author}{\bibfnamefont{M.}~\bibnamefont{Nakamura}} \bibnamefont{and}
  \bibinfo{author}{\bibfnamefont{D.~L.} \bibnamefont{Meier}},
  \bibinfo{journal}{Astrophys. J.} \textbf{\bibinfo{volume}{617}},
  \bibinfo{pages}{123} (\bibinfo{year}{2004}).

\bibitem[{\citenamefont{Nakamura et~al.}(2006)\citenamefont{Nakamura, Li, and
  Li}}]{nll06}
\bibinfo{author}{\bibfnamefont{M.}~\bibnamefont{Nakamura}},
  \bibinfo{author}{\bibfnamefont{H.}~\bibnamefont{Li}}, \bibnamefont{and}
  \bibinfo{author}{\bibfnamefont{S.}~\bibnamefont{Li}},
  \bibinfo{journal}{Astrophys. J.} \textbf{\bibinfo{volume}{652}},
  \bibinfo{pages}{1059} (\bibinfo{year}{2006}).

\bibitem[{\citenamefont{Nakamura et~al.}(2007)\citenamefont{Nakamura, Li, and
  Li}}]{nll07}
\bibinfo{author}{\bibfnamefont{M.}~\bibnamefont{Nakamura}},
  \bibinfo{author}{\bibfnamefont{H.}~\bibnamefont{Li}}, \bibnamefont{and}
  \bibinfo{author}{\bibfnamefont{S.}~\bibnamefont{Li}},
  \bibinfo{journal}{Astrophys. J.} \textbf{\bibinfo{volume}{656}},
  \bibinfo{pages}{721} (\bibinfo{year}{2007}).

\bibitem[{\citenamefont{Lynn et~al.}(2007)\citenamefont{Lynn, Zhang, Hsu, Li,
  Liu, Gilmore, and Watts}}]{lzh07}
\bibinfo{author}{\bibfnamefont{A.~G.} \bibnamefont{Lynn}},
  \bibinfo{author}{\bibfnamefont{Y.}~\bibnamefont{Zhang}},
  \bibinfo{author}{\bibfnamefont{S.~C.} \bibnamefont{Hsu}},
  \bibinfo{author}{\bibfnamefont{H.}~\bibnamefont{Li}},
  \bibinfo{author}{\bibfnamefont{W.}~\bibnamefont{Liu}},
  \bibinfo{author}{\bibfnamefont{M.}~\bibnamefont{Gilmore}}, \bibnamefont{and}
  \bibinfo{author}{\bibfnamefont{C.}~\bibnamefont{Watts}}, in
  \emph{\bibinfo{booktitle}{Bull. Amer. Phys. Soc.}} (\bibinfo{year}{2007}),
  vol.~\bibinfo{volume}{52}, p.~\bibinfo{pages}{53}.

\bibitem[{\citenamefont{Lynn et~al.}(2008)\citenamefont{Lynn, Gilmore, and
  Watts}}]{lgw08}
\bibinfo{author}{\bibfnamefont{A.~G.} \bibnamefont{Lynn}},
  \bibinfo{author}{\bibfnamefont{M.}~\bibnamefont{Gilmore}}, \bibnamefont{and}
  \bibinfo{author}{\bibfnamefont{C.}~\bibnamefont{Watts}},
  \bibinfo{journal}{Review of Scientific Instruments}  (\bibinfo{year}{2008}),
  \bibinfo{note}{in press}.

\bibitem[{\citenamefont{Li and Li}(2003)}]{ll03}
\bibinfo{author}{\bibfnamefont{H.}~\bibnamefont{Li}} \bibnamefont{and}
  \bibinfo{author}{\bibfnamefont{S.}~\bibnamefont{Li}},
  \bibinfo{type}{Technical Report LA-UR-03-8935}, \bibinfo{institution}{Los
  Alamos National Laboratory} (\bibinfo{year}{2003}).

\bibitem[{\citenamefont{Geddess et~al.}(1998)\citenamefont{Geddess, Kornack,
  and Brown}}]{gkb98}
\bibinfo{author}{\bibfnamefont{C.~G.~R.} \bibnamefont{Geddess}},
  \bibinfo{author}{\bibfnamefont{T.~W.} \bibnamefont{Kornack}},
  \bibnamefont{and} \bibinfo{author}{\bibfnamefont{M.~R.} \bibnamefont{Brown}},
  \bibinfo{journal}{Phys. Plasmas} \textbf{\bibinfo{volume}{5}},
  \bibinfo{pages}{1027} (\bibinfo{year}{1998}).

\bibitem[{\citenamefont{Yee and Bellan}(2000)}]{yb00}
\bibinfo{author}{\bibfnamefont{J.}~\bibnamefont{Yee}} \bibnamefont{and}
  \bibinfo{author}{\bibfnamefont{P.~M.} \bibnamefont{Bellan}},
  \bibinfo{journal}{Phys. Plasmas} \textbf{\bibinfo{volume}{7}},
  \bibinfo{pages}{3625} (\bibinfo{year}{2000}).

\bibitem[{\citenamefont{Hsu and Bellan}(2003)}]{hb03}
\bibinfo{author}{\bibfnamefont{S.~C.} \bibnamefont{Hsu}} \bibnamefont{and}
  \bibinfo{author}{\bibfnamefont{P.~M.} \bibnamefont{Bellan}},
  \bibinfo{journal}{Phys. Rev. Lett.} \textbf{\bibinfo{volume}{90}},
  \bibinfo{pages}{215002} (\bibinfo{year}{2003}).

\bibitem[{\citenamefont{Hsu and Bellan}(2005)}]{hb05}
\bibinfo{author}{\bibfnamefont{S.~C.} \bibnamefont{Hsu}} \bibnamefont{and}
  \bibinfo{author}{\bibfnamefont{P.~M.} \bibnamefont{Bellan}},
  \bibinfo{journal}{Phys. Plasmas} \textbf{\bibinfo{volume}{12}},
  \bibinfo{pages}{032103} (\bibinfo{year}{2005}).

\bibitem[{\citenamefont{Shafranov}(1957)}]{sv57}
\bibinfo{author}{\bibfnamefont{V.~D.} \bibnamefont{Shafranov}},
  \bibinfo{journal}{Soviet Phys. JETP} \textbf{\bibinfo{volume}{6}},
  \bibinfo{pages}{545} (\bibinfo{year}{1957}).

\bibitem[{\citenamefont{Kruskal et~al.}(1958)\citenamefont{Kruskal, Johnson,
  Gottlieb, and Goldman}}]{kjgg58}
\bibinfo{author}{\bibfnamefont{M.~D.} \bibnamefont{Kruskal}},
  \bibinfo{author}{\bibfnamefont{J.~L.} \bibnamefont{Johnson}},
  \bibinfo{author}{\bibfnamefont{M.~B.} \bibnamefont{Gottlieb}},
  \bibnamefont{and} \bibinfo{author}{\bibfnamefont{L.~M.}
  \bibnamefont{Goldman}}, \bibinfo{journal}{Phys. Fluids}
  \textbf{\bibinfo{volume}{1}}, \bibinfo{pages}{421} (\bibinfo{year}{1958}).

\bibitem[{\citenamefont{Hood and Priest}(1979)}]{hp79}
\bibinfo{author}{\bibfnamefont{A.~W.} \bibnamefont{Hood}} \bibnamefont{and}
  \bibinfo{author}{\bibfnamefont{E.~R.} \bibnamefont{Priest}},
  \bibinfo{journal}{Sol. Phys.} \textbf{\bibinfo{volume}{64}},
  \bibinfo{pages}{303} (\bibinfo{year}{1979}).

\bibitem[{\citenamefont{Einaudi and Van~Hoven}(1983)}]{eh83}
\bibinfo{author}{\bibfnamefont{G.}~\bibnamefont{Einaudi}} \bibnamefont{and}
  \bibinfo{author}{\bibfnamefont{G.}~\bibnamefont{Van~Hoven}},
  \bibinfo{journal}{Sol.Phys.} \textbf{\bibinfo{volume}{88}},
  \bibinfo{pages}{163} (\bibinfo{year}{1983}).

\bibitem[{\citenamefont{Furno et~al.}(2006)\citenamefont{Furno, Intrator,
  Ryutov, Abbate, Madziwa-Nussinov, Light, Dorf, and Lapenta}}]{fir06}
\bibinfo{author}{\bibfnamefont{I.}~\bibnamefont{Furno}},
  \bibinfo{author}{\bibfnamefont{T.~P.} \bibnamefont{Intrator}},
  \bibinfo{author}{\bibfnamefont{D.~D.} \bibnamefont{Ryutov}},
  \bibinfo{author}{\bibfnamefont{S.}~\bibnamefont{Abbate}},
  \bibinfo{author}{\bibfnamefont{T.}~\bibnamefont{Madziwa-Nussinov}},
  \bibinfo{author}{\bibfnamefont{A.}~\bibnamefont{Light}},
  \bibinfo{author}{\bibfnamefont{L.}~\bibnamefont{Dorf}}, \bibnamefont{and}
  \bibinfo{author}{\bibfnamefont{G.}~\bibnamefont{Lapenta}},
  \bibinfo{journal}{Phys. Rev. Lett.} \textbf{\bibinfo{volume}{97}},
  \bibinfo{pages}{015002} (\bibinfo{year}{2006}).

\bibitem[{\citenamefont{Tomimatsu et~al.}(2001)\citenamefont{Tomimatsu,
  Matsuoka, and Takahashi}}]{tmt01}
\bibinfo{author}{\bibfnamefont{A.}~\bibnamefont{Tomimatsu}},
  \bibinfo{author}{\bibfnamefont{T.}~\bibnamefont{Matsuoka}}, \bibnamefont{and}
  \bibinfo{author}{\bibfnamefont{M.}~\bibnamefont{Takahashi}},
  \bibinfo{journal}{Phys. Rev. D} \textbf{\bibinfo{volume}{64}},
  \bibinfo{pages}{123003} (\bibinfo{year}{2001}).

\bibitem[{\citenamefont{Meier et~al.}(2001)\citenamefont{Meier, Koide, and
  Uchida}}]{mku01}
\bibinfo{author}{\bibfnamefont{D.~L.} \bibnamefont{Meier}},
  \bibinfo{author}{\bibfnamefont{S.}~\bibnamefont{Koide}}, \bibnamefont{and}
  \bibinfo{author}{\bibfnamefont{Y.}~\bibnamefont{Uchida}},
  \bibinfo{journal}{Science} \textbf{\bibinfo{volume}{291}},
  \bibinfo{pages}{84} (\bibinfo{year}{2001}).

\bibitem[{\citenamefont{Balbus and Hawley}(1998)}]{bh98}
\bibinfo{author}{\bibfnamefont{S.~A.} \bibnamefont{Balbus}} \bibnamefont{and}
  \bibinfo{author}{\bibfnamefont{J.~F.} \bibnamefont{Hawley}},
  \bibinfo{journal}{Rev. Mod. Phys.} \textbf{\bibinfo{volume}{70}},
  \bibinfo{pages}{1} (\bibinfo{year}{1998}).

\end{thebibliography}

\clearpage
%

%

%


\clearpage

\clearpage

\clearpage

\clearpage

%

%

\begin{table}[!htp]
\begin{tabular}{c|c|c|c}

\hline 
Physical Quantities &
Description &
Normalized Units &
Typical Values \tabularnewline
\hline
\hline 
$R$&
Length &
$R_0$&
$3\;\unit{cm}$\tabularnewline
\hline 
$\vec{V}$&
Velocity Field&
$C_{s0}$&
$2.0\times10^{5}\;\unit{cm\; s^{-1}}$\tabularnewline
\hline 
$t$&
Time&
$R_0/C_{s0}$&
$1.50\times10^{-5}\;\unit{s}$\tabularnewline
\hline 
$n$&
Number Density&
$n_0$&
$10^{12}\;\unit{cm^{-3}}$\tabularnewline
\hline
$\rho$&
Density&
$\rho_0=n_0M_0$\footnote{$M_0$ is the atomic mass. In the experiment it is the argon atomic mass $M_0=6.67\times10^{-23}\;\unit{g}$.}&
$6.67\times10^{-11}\;\unit{g\; cm^{-3}}$\tabularnewline
\hline
$p$&
Pressure&
$\rho_0 C_{s0}^2$&
$2.67\;\unit{dyn\; cm^{-2}}$\tabularnewline
\hline
$\vec{B}$&
Magnetic Field&
$(4\pi\rho_0 C_{s0}^2)^{1/2}$&
$5.79\;\unit{Gauss}$\tabularnewline
\hline
$T$&
Temperature&
$T_0$&
$1\;\unit{ev}$\tabularnewline
\hline
$F_{\rm Lorentz}$&
Lorentz Force&
$F_0$&
$11.2\;\unit{dyne}$\tabularnewline
\hline
\end{tabular}
\caption{\label{normalization}Physical quantities, normalization constants, and values.}
\end{table}

\begin{table}[!htp]
\begin{center}
\begin{tabular}{c|c|c}

\hline 
$V_{\rm inj}/V_{A,0}$&
 Slow-Mode Wave Front Speed &
 Perpendicular Shock Speed\tabularnewline
\hline
\hline 
0.0035&
3.6&
7.2\tabularnewline
\hline 
0.18&
3.6&
7.2\tabularnewline
\hline 
0.5&
3.6&
7.2\tabularnewline
\hline 
1.0&
3.6&
7.2\tabularnewline\hline 
1.8&
3.6&
10.8\tabularnewline
\hline 
3.5&
10.8&
18\tabularnewline
\hline
35.4&
160&
180\tabularnewline
\hline

\end{tabular}
\end{center}
\caption{\label{speed}Shock speed and wavefront speed vs.\ injection velocity with $\alpha=\sqrt{10}$. All velocities are normalized to the sound speed $C_{s0}$. $V_{A,0}=4.87$.}
\end{table}

\begin{table}[!htp]
\begin{center}
\begin{tabular}{c|c|c|c|c|c|c|c|c}

\hline 
&
$\rho$ &
$p$ &
$V_x$ &
$V_y$ &
$V_z$ &
$B_x$ &
$B_y$ &
$B_z$ \tabularnewline
\hline
\hline 
Upstream Region&
0.87&
2.44&
0&
0&
-7.2&
2.87&
0&
0\tabularnewline\hline 
Downstream Region Simulation&
1.35&
9.6&
0&
0.29&
-3.58&
5.79&
0&
0.14\tabularnewline
\hline 
Downstream Region Calculation&
1.81&
12.12&
0&
0&
-3.45&
5.99&
0&
0\tabularnewline
\hline 

\end{tabular}
\end{center}
\caption{\label{jump} Values of physical quantities across the MHD shock. All velocities are in the
shock rest frame, and all magnetic field values are normalized by $\sqrt{4\pi}$.}
\end{table}

\begin{table}[!htp]
\begin{center}
\begin{tabular}{c|c|c|c|c}
\hline 
&
$t=0$ &
$t=0.25$&
$t=0.5$& 
$t=1.0$\tabularnewline
\hline
\hline 
$\psi_t=\int B_y dS$&
134.9&
96.9&
88.5&
64.3\tabularnewline
\hline 
\hline 
\end{tabular}
\end{center}
\caption{~Decay of the net toroidal magnetic flux $\psi_t=\int B_y dS$, where only positive $B_y$ is selected. 
\label{toroidal}}
\end{table}

\end{document}